 \documentclass[11pt]{article}

\usepackage{amsmath,amsfonts,amssymb}

 \usepackage[numbers,sort&compress]{natbib}

\allowdisplaybreaks[1]
 
\usepackage{hyperref}
\hypersetup{
   colorlinks   =  true,
    linkcolor    = blue,
    citecolor    = red,
     urlcolor	=blue,     
	urlbordercolor = {1 1 0}
}

\newcommand\be{\begin{equation}}
\newcommand\ee{\end{equation}}
\newcommand\bea{\begin{eqnarray}}
\newcommand\eea{\end{eqnarray}}

\usepackage{lmodern}
\usepackage[english]{babel}
\usepackage[utf8]{inputenc}
\usepackage[T1]{fontenc}

\usepackage{doi}
\usepackage{pgf,tikz,pgfplots}
\usepackage{tikz-cd}
\usetikzlibrary{babel}
\usepackage{svg}
\pgfplotsset{compat=1.15}
\usetikzlibrary{matrix,arrows}
\usepackage{graphicx}
\usepackage{adjustbox}
\usepackage{subcaption,booktabs}
\graphicspath{{graficos/}}

\usepackage{amsmath}
\usepackage{amsthm}
\usepackage{physics}
\usepackage{amssymb}
\usepackage{stmaryrd}
\usepackage{bm}
\usepackage{framed}

\theoremstyle{definition}

\theoremstyle{plain}

\theoremstyle{remark}

\numberwithin{equation}{section}

\begin{document}
\
  \vskip0.5cm

  \begin{center}     
 \noindent
 {\Large \bf  
Soliton hierarchies associated with Lie algebra $\mathfrak{sp}(6)$} 
 
   \end{center}

\medskip

\begin{center}

{\sc  Yanhui Bi$^{1}$, Yuqi Ruan$^{2}$ , Bo Yuan$^{2}$ and Tao Zhang$^{3}$}

\end{center}

\medskip

 \noindent
$^1$ Center for Mathematical Sciences, College of Mathematics and Information Science,
Nanchang Hangkong University, Nanchang, China

\noindent 
$^2$ College of Mathematics and Information Science,
Nanchang Hangkong University, Nanchang, China

\noindent
$^3$ School of Mathematics and Statistics,
	Henan Normal University, Xinxiang, China

 \medskip
 
\noindent  E-mail: {\small
 \href{mailto:biyanhui0523@163.com}{\texttt{biyanhui0523@163.com}}, \href{mailto:ruanyuqi1023@163.com}{\texttt{ruanyuqi1023@163.com}}, \href{mailto:yuanbo010806@163.com}{\texttt{yuanbo010806@163.com}},
 \href{mailto:zhangtao@htu.edu.cn}{\texttt{zhangtao@htu.edu.cn}},
}
 
\medskip

\begin{abstract}
\noindent

In this paper, by selecting appropriate spectral matrices within the loop algebra of symplectic Lie algebra $\mathfrak{sp}(6)$, we construct two distinct classes of integrable soliton hierarchies. Then, by employing the Tu scheme and trace identity, we derive the Hamiltonian structures of the aforementioned two classes of integrable systems. From these two classes of integrable soliton hierarchies, we select one particular hierarchy and employ the Kronecker product to construct an integrable coupling system.
\end{abstract}
\medskip
\medskip


 
\smallskip
\noindent
\textbf{PACS 2010 codes}:  {02.30.Hq, 02.20.Sv, 45.20.Jj, 45.10.Na}\\


\section{Introduction}
Soliton theory and integrable systems represent a significant branch of applied mathematics and mathematical physics, characterized by a rich variety of content and research methodologies. On the one hand, soliton equations may be derived through geometric approaches or by employing geometric tools; on the other hand, the algebraic properties of integrable systems can be effectively studied using Lie algebra theory. In recent years, rapid developments in mathematical physics and computer algebra have contributed to substantial advances in the study of soliton theory and integrable systems.

In mathematical physics, studying the generation of integrable systems and uncovering their algebraic-geometric properties is an important research direction. There exist multiple approaches to constructing integrable systems and analyzing their properties \cite{ref17}. We can further construct their integrable coupling systems, as demonstrated in {}\cite{ref3,ref7,ref8,ref10,ref12,ref15,ref20}. In {}\cite{ref13}, Gui-Zhang Tu proposed an effective approach for generating integrable systems and deriving their Hamiltonian structures. Wen-Xiu Ma later referred to this method as the Tu Scheme. Researchers have since used the Tu Scheme to generate numerous integrable systems and obtain their corresponding Hamiltonian structures \cite{ref6}.

The trace identity is a powerful tool for constructing the Hamiltonian structure of integrable systems \cite{ref9,ref13,ref14}. We first consider an isospectral problem
\begin{align*}
	\begin{cases}
		\phi_{x}=U\phi,\\
		\phi_{t}=V\phi,
	\end{cases}
\end{align*}
where $\phi$ is an $n$-dimensional vector function of $x$ and $t$, $U$ and $V$ are $n\times n$ matrices whose entries depend on the spectral parameter and an $m$-dimensional vector function $u(x,t)$ and its partial derivatives of the variables $x$ and $t$. For the above two equations to be simultaneously solvable, $\phi$ must satisfy the compatibility condition $\phi_{xt}=\phi_{tx}$. This leads to the zero-curvature equation \cite{ref16}: $U_{t}-V_{x}+[U,V]=0$, and stationary zero curvature representation: $V_{x}=[U,V]$. \\

By selecting appropriate spectral matrices $U$ and $V$, one can derive  a hierarchy of soliton equations. Let $U=\{U_{ij}(n)\}_{6\times6},V=\sum_{k\geq0}\{V_{ij}(-k)\}_{6\times6}\in\widetilde{\mathfrak{sp}(6)}$, from the stationary zero curvature equation $V_{x}=[U,V]$, we can derive the solution for $V_{x}$. We then consider
\begin{align*}
	V^{n}=\lambda^{n}V=\sum_{k\geq0}\{V_{ij}(n-k)\}_{6\times6},\ V_{+}^{n}=\sum_{k\geq0}^{n}\{V_{ij}(n-k)\}_{6\times6},\ V_{-}^{n}=V^{n}-V_{+}^{n}.
\end{align*}
From the zero-curvature equation $U_{t}-V_{x}+[U,V]=0$, we obtain a hierarchy of soliton equations
\begin{align*}
	u_{t}=J\frac{\delta H_{n}}{\delta u}.
\end{align*}

To put the hierarchy in its Hamiltonian form we apply the trace identity \cite{ref9,ref13,ref14}
\begin{align*}
	(\frac{\partial}{\partial u_{i}})\langle V,\frac{\partial U}{\partial \lambda}\rangle=(\lambda^{-\tau}(\frac{\partial}{\partial\lambda})\lambda^{\tau})\langle V,\frac{\partial U}{\partial u_{i}}\rangle,
\end{align*}
where $\langle x,y\rangle=tr(xy)$, $\tau=\frac{\lambda}{2}ln|tr(V^{2})|$.
It is said that the hierarchy have the Hamiltonian structure, and it is easy to verity the hierarchy is Liouville integrable \cite{ref6,ref13,ref14}.\\

In {}\cite{ref3}, F. Guo et al. employed the Tu scheme to systematically derive both the multicomponent KN hierarchy and its integrable coupling system. In our work, after obtaining two integrable hierarchies associated with $\mathfrak{sp}(6)$ through the Tu scheme, we further constructed an integrable coupling system for one of them using the Kronecker product method \cite{ref8}.

In {}\cite{ref1}, H.H. Dong et al. introduced the application of the Tu scheme to loop algebras. In {}\cite{ref2}, B.L. Feng et al. derived two new families of integrable equations based on a subalgebra of loop algebras and its extended loop algebra structure, while W.D. Zhao et al. in {}\cite{ref21} carried out similar studies on different Lie algebra and its loop algebra. In {}\cite{ref11}, W.X. Ma constructed an integrable soliton hierarchy associated with $\mathfrak{so}(3,\mathbb{R})$. In {}\cite{ref18}, H.Y. Wei et al. constructed a new Lie algebra by using the basis of $\mathfrak{sl}(2)$ as matrix blocks, and subsequently built integrable systems over this resulting Lie algebra. In {}\cite{ref4}, B. He et al. constructed integrable hierarchies by selecting spectral matrices based on the Lie algebras $\mathfrak{sp}(4)$ and $\mathfrak{so}(5)$. The higher-dimensional case $\mathfrak{sp}(6)$ is the Lie algebra comprising all $6\times 6$ complex matrices $X$ such that $X^{t}J+JX=0$, i.e.,
\begin{align*}
	\mathfrak{sp}(6)=\{X\in\mathfrak{gl}(6,\mathbb{C})|X^{t}J+JX=0\},
\end{align*}
where is the standard symplectic form given by
\begin{align*}
	J=\left(\begin{array}{cc}
		0&I_{3}\\
		-I_{3}&0
	\end{array}\right),
\end{align*}
and $I_{3}$ denotes the $3\times 3$ identity matrix \cite{ref22,ref23}. It is straightforward to verify that $\mathfrak{sp}(6)$ is a 21-dimensional Lie algebra. In {}\cite{ref22}, O. Carballal et al. obtained new classes of Lie-Hamilton systems from the six-dimensional fundamental representation of the symplectic Lie algebra $\mathfrak{sp}(6,\mathbb{R})$. In {}\cite{ref23}, A. Molev constructed a basis for each finite-dimensional irreducible representation of the symplectic Lie algebra $\mathfrak{sp}(2n)$. Since $\mathfrak{sp}(6)$ is a semisimple Lie algebra (admitting no non-zero solvable ideals), we naturally extend our investigation to its loop algebra formulation to construct integrable systems.
The loop algebra associated with $\mathfrak{sp}(6)$ is defined as follows
\begin{align*}
	\widetilde{\mathfrak{sp}(6)}=span\{E_{i}(n)\}_{i=1}^{21},
\end{align*}
where $\{E_{i}\}_{i=1}^{21}$ is a basis of $\mathfrak{sp}(6)$.
Using the aforementioned method, numerous integrable soliton families can be constructed. However, the study of constructing integrable soliton hierarchies on the symplectic Lie algebra $\mathfrak{sp}(6)$ remains unexplored. In this paper we will fill this gap by constructing an integrable system on the loop algebra of $\mathfrak{sp}(6)$.

The paper is organized as follows. In Section \ref{Sec:2-1}, we construct a family of integrable equations through appropriate choices of $U_{0}, V_{0}\in\widetilde{\mathfrak{sp}(6)}$. By selecting $U_{1}\in\widetilde{\mathfrak{sp}(6)}$ and taking $V_{0}$ as defined in Section \ref{Sec:2-1}, we constructed another integrable system in Section \ref{Sec:2-2}. In Section \ref{Sec:3}, we choose $U_{2},V_{2}\in\widetilde{\mathfrak{sp}(6)}$,
and construct $U_{3}$ from $U_{0}$ and $U_{2}$, and $V_{3}$ from $V_{0}$ and $V_{2}$ by Kronecker product. We further construct an integrable coupling system associated with the integrable system presented in Section \ref{Sec:2-1}.

%

\section{ Two Soliton Hierarchies Associated with $\mathfrak{sp}(6)$}\label{Sec:2}

In this section, we construct two integrable soliton hierarchies by selecting different spectral matrices in the loop algebra of $\mathfrak{sp}(6)$ and utilizing the Tu scheme. Building upon this foundation, we choose one of them and employ the Kronecker product to construct an integrable coupling system for it.

\subsection{ The First Soliton Hierarchy Associated with $\mathfrak{sp}(6)$.\\}\label{Sec:2-1}
The compact real form $\mathfrak{sp}(6)$ of complex symplectic Lie algebra $\mathfrak{sp}(6,\mathbb{C})$ is defined as
\begin{align*}
	\mathfrak{sp}(6)=\{X\in\mathfrak{gl}(6,\mathbb{C})|X^{t}J+JX=0 \},
\end{align*}
where $J=\left(
\begin{array}{cccc}
	0&I_{3}\\
	-I_{3}&0\\
\end{array}\right)$, and $I_{3}$ is the $3\times3$ identity matrix. We can obtain the bases of Lie algebra $\mathfrak{sp}(6)$\cite{ref23}

\begin{equation}
	\begin{aligned}
		&E_{1}=e_{11}-e_{44}, E_{2}=e_{22}-e_{55}, E_{3}=e_{33}-e_{66}, E_{4}=e_{12}-e_{54}, E_{5}=e_{21}-e_{45},\\
		& E_{6}=e_{13}-e_{64}, E_{7}=e_{31}-e_{46},E_{8}=e_{23}-e_{65}, E_{9}=e_{32}-e_{56}, E_{10}=e_{16}+e_{34},\\
		&  E_{11}=e_{43}+e_{61}, E_{12}=e_{25}, E_{13}=e_{52},E_{14}=e_{14},E_{15}=e_{41}, E_{16}=e_{15}+e_{24}, \\
		&E_{17}=e_{42}+e_{15}, E_{18}=e_{26}+e_{35}, E_{19}=e_{53}+e_{62}, E_{20}=e_{36}, E_{21}=e_{63},
	\end{aligned}\label{eq:2.1}
\end{equation}

where $e_{ij}$ is a $6\times 6$ matrix with $1$ in the $(i, j)$-th position and $0$ elsewhere. The subspace $span\{E_{1},E_{2},E_{3}\}$ is a Cartan subalgebra of $\mathfrak{sp}(6)$, and thus the commutator $[E_{i},E_{j}]=0$ for all $i,j\in\{1,2,3\}$. The remaining structure constants are given as follows
\begin{align*}
	&[E_{1},E_{4}] = E_{4},\
	[E_{1},E_{5}] = -E_{5},\
	[E_{1},E_{6}] = E_{6},\
	[E_{1},E_{7}] = -E_{7},\\
	&[E_{1},E_{10}] = E_{10},\
	[E_{1},E_{11}] = -E_{11},\
	[E_{1},E_{14}] = 2E_{14},\
	[E_{1},E_{15}] = -2E_{15},\\
	&[E_{1},E_{16}] = E_{16},\
	[E_{1},E_{17}] = -E_{17},\
	[E_{2},E_{4}] = -E_{4},\
	[E_{2},E_{5}] = E_{5},\\
	&[E_{2},E_{8}] = E_{8},\
	[E_{2},E_{9}] = -E_{9},\
	[E_{2},E_{12}] = 2E_{12},\
	[E_{2},E_{13}] = -2E_{13},\\
	&[E_{2},E_{16}] = E_{16},\
	[E_{2},E_{17}] = -E_{17},\
	[E_{2},E_{18}] = E_{18},\
	[E_{2},E_{19}] = -E_{19},\\
	&[E_{3},E_{6}] = -E_{6},\
	[E_{3},E_{7}] = E_{7},\
	[E_{3},E_{8}] = -E_{8},\
	[E_{3},E_{9}] = E_{9},\\
	&[E_{3},E_{10}] = E_{10},\
	[E_{3},E_{11}] = -E_{11},\
	[E_{3},E_{18}] = E_{18},\
	[E_{3},E_{19}] = -E_{19},\\
	&[E_{3},E_{20}] = 2E_{20},\
	[E_{3},E_{21}] = -2E_{21},\
	[E_{4},E_{5}] = E_{1}-E_{2},\
	[E_{4},E_{7}] = -E_{9},\\
	&[E_{4},E_{8}] = E_{6},\
	[E_{4},E_{11}] = -E_{19},\
	[E_{4},E_{12}] = E_{16},\
	[E_{4},E_{15}] = -E_{17},\\
	&[E_{4},E_{16}] = 2E_{14},\
	[E_{4},E_{17}] = -2E_{13},\
	[E_{4},E_{18}] = E_{10},\
	[E_{5},E_{6}] = E_{8},\\
	&[E_{5},E_{9}] = -E_{7},\
	[E_{5},E_{10}] = E_{18},\
	[E_{5},E_{13}] = -E_{17},\
	[E_{5},E_{14}] = E_{16},\\
	&[E_{5},E_{16}] = 2E_{12},\
	[E_{5},E_{17}] = -2E_{15},\
	[E_{5},E_{19}] = -E_{11},\
	[E_{6},E_{7}] = E_{1}-E_{3},\\
	&[E_{6},E_{9}] = E_{4},\
	[E_{6},E_{10}] = 2E_{14},\
	[E_{6},E_{11}] = -2E_{21},\
	[E_{6},E_{15}] = -E_{11},\\
	&[E_{6},E_{17}] = -E_{19},\
	[E_{6},E_{18}] = E_{16},\
	[E_{6},E_{20}] = E_{10},\
	[E_{7},E_{8}] = -E_{5},\\
	&[E_{7},E_{10}] = 2E_{20},\
	[E_{7},E_{11}] = -2E_{15},\
	[E_{7},E_{14}] = E_{10},\
	[E_{7},E_{16}] = E_{18},\\
	&[E_{7},E_{19}] = -E_{17},\
	[E_{7},E_{21}] = -E_{11},\
	[E_{8},E_{9}] = E_{2}-E_{3},\
	[E_{8},E_{10}] = E_{16},\\
	&[E_{8},E_{13}] = -E_{19},\
	[E_{8},E_{17}] = -E_{11},\
	[E_{8},E_{18}] = 2E_{12},\
	[E_{8},E_{19}] = -2E_{21},\\
	&[E_{8},E_{20}] = E_{18},\
	[E_{9},E_{11}] = -E_{17},\
	[E_{9},E_{12}] = E_{18},\
	[E_{9},E_{16}] = E_{10},\\
	&[E_{9},E_{18}] = 2E_{20},\
	[E_{9},E_{19}] = -2E_{13},\
	[E_{9},E_{21}] = -E_{19},\
	[E_{10},E_{11}] = E_{1}+ E_{3},\\
	&[E_{10},E_{15}] = E_{7},\
	[E_{10},E_{17}] = E_{9},\
	[E_{10},E_{19}] = E_{4},\
	[E_{10},E_{21}] = E_{6},\\
	&[E_{11},E_{14}] = -E_{6},\
	[E_{11},E_{16}] = -E_{8},\
	[E_{11},E_{18}] = -E_{5},\
	[E_{11},E_{20}] = -E_{7},\\
	&[E_{12},E_{13}] = E_{2},\
	[E_{12},E_{17}] = E_{5},\
	[E_{12},E_{19}] = E_{8},\
	[E_{13},E_{16}] = -E_{4},\\
	&[E_{13},E_{18}] = -E_{9},\
	[E_{14},E_{15}] = E_{1},\
	[E_{14},E_{17}] = E_{4},\
	[E_{15},E_{16}] = -E_{5},\\
	&[E_{16},E_{17}] = E_{1}+ E_{2},\
	[E_{16},E_{19}] = E_{6},\
	[E_{17},E_{18}] = -E_{7},\
	[E_{18},E_{19}] = E_{2}+ E_{3},\\
	&[E_{18},E_{21}] = E_{8},\
	[E_{19},E_{20}] = -E_{9},\
	[E_{20},E_{21}] = E_{3}.
\end{align*}
It is straightforward to verify that $\mathfrak{sp}(6)$ admits no nonzero solvable ideals, and hence it is a semisimple Lie algebra.
We consider the loop algebra of $\mathfrak{sp}(6)$
\begin{align*}
	\widetilde{\mathfrak{sp}(6)}=span\{E_{1}(n),\dots,E_{21}(n)\},
\end{align*}
where $E_{i}(n)=E_{i}\lambda^{n},\ [E_{i}(m),E_{j}(n)]=[E_{i},E_{j}]\lambda^{m+n}.$

We are going to construct a soliton hierarchy from the loop algebra $\widetilde{\mathfrak{sp}(6)}$. Consider an isospectral problem
\begin{align*}
	\begin{cases}
		\phi_{x} = U_{0}\phi, \\
		\phi_{t} = V_{0}\phi, \quad \lambda_{t} = 0.
	\end{cases}
\end{align*}

Let $U_{0}, V_{0}\in \widetilde{\mathfrak{sp}(6)}$ be given as follows
\begin{align*}
	U_{0}=&E_{1}(1)+E_{2}(1)+E_{3}(1)+u_{1}E_{10}(0)+u_{2}E_{11}(0)+u_{3}E_{12}(0)+u_{4}E_{13}(0)+u_{5}E_{14}(0)\\
	&+u_{6}E_{15}(0)+u_{7}E_{16}(0)+u_{8}E_{17}(0)+u_{9}E_{18}(0)+u_{10}E_{19}(0)+u_{11}E_{20}(0)+u_{12}E_{21}(0)\\
	=&(\lambda,\lambda,\lambda,0,0,0,0,0,0,u_{1},u_{2},u_{3},u_{4},u_{5},u_{6},u_{7},u_{8},u_{9},u_{10},u_{11},u_{12})^{t},
\end{align*}
i.e.
\begin{align}
	U_{0}=\left(
	\begin{array}{cccccc}
		\lambda&0&0&u_{5}&u_{7}&u_{1}\\
		0&\lambda&0&u_{7}&u_{3}&u_{9}\\
		0&0&\lambda&u_{1}&u_{9}&u_{11}\\
		u_{6}&u_{8}&u_{2}&-\lambda&0&0\\
		u_{8}&u_{4}&u_{10}&0&-\lambda&0\\
		u_{2}&u_{10}&u_{12}&0&0&-\lambda
	\end{array}\right)\label{eq:2.2}
\end{align}
and
\begin{align*}
	V_{0}=&aE_{1}(0)+bE_{2}(0)+cE_{3}(0)+dE_{4}(0)+eE_{5}(0)+fE_{6}(0)+gE_{7}(0)+hE_{8}(0)+jE_{9}(0)\\
	&+kE_{10}(0)+lE_{11}(0)+mE_{12}(0)+oE_{13}(0)+pE_{14}(0)+qE_{15}(0)+rE_{16}(0)+sE_{17}(0)\\
	&+tE_{18}(0)+uE_{19}(0)+vE_{20}(0)+wE_{21}(0)\\
	=&(a,b,c,d,e,f,g,h,j,k,l,m,o,p,q,r,s,t,u,v,w)^{t},
\end{align*}
where $a=\sum_{i\geq0}a_{i}\lambda^{-i},\dots,w=\sum_{i\geq0}w_{i}\lambda^{-i},$
i.e.
\begin{align}
	V_{0}=\left(
	\begin{array}{cccccc}
		a&d&f&p&r&k\\
		e&b&h&r&m&t\\
		g&j&c&k&t&v\\
		q&s&l&-a&-e&-g\\
		s&o&u&-d&-b&-j\\
		l&u&w&-f&-h&-c\\
	\end{array}\right)= \sum_{i\geq0}\left(
	\begin{array}{cccccc}
		a_{i}&d_{i}&f_{i}&p_{i}&r_{i}&k_{i}\\
		e_{i}&b_{i}&h_{i}&r_{i}&m_{i}&t_{i}\\
		g_{i}&j_{i}&c_{i}&k_{i}&t_{i}&v_{i}\\
		q_{i}&s_{i}&l_{i}&-a_{i}&-e_{i}&-g_{i}\\
		s_{i}&o_{i}&u_{i}&-d_{i}&-b_{i}&-j_{i}\\
		l_{i}&u_{i}&w_{i}&-f_{i}&-h_{i}&-c_{i}\\
	\end{array}\right)\lambda^{-i}\label{eq:2.3}
\end{align}
The stationary zero curvature representation $V_{0,x}=[U_{0},V_{0}]$ gives
\begin{align}
	\begin{cases}
		a_{x}=u_{1}l - u_{2}k + u_{5}q - u_{6}p  + u_{7}s - u_{8}r,\\
		b_{x}=u_{3}o - u_{4}m + u_{7}s- u_{8}r+ u_{9}u - u_{10}t ,\\
		c_{x}=u_{1}l - u_{2}k + u_{9}u- u_{10}t +u_{11}w - u_{12}v  ,\\
		d_{x}=u_{1}u- u_{4}r+ u_{5}s +u_{7}o - u_{8}p- u_{10}k ,\\
		e_{x}=- u_{2}t+ u_{3}s- u_{6}r + u_{7}q - u_{8}m +u_{9}l  ,\\
		f_{x}=u_{1}w- u_{2}p+u_{5}l+ u_{7}u-u_{10}r-u_{12}k
		,\\
		g_{x}=u_{1}q-u_{2}v-u_{6}k-u_{8}t+u_{9}s+u_{11}l
		,\\
		h_{x}=- u_{2}r+u_{3}u+u_{7}l+u_{9}w-u_{10}m-u_{12}t
		,\\
		j_{x}=u_{1}s- u_{4}t- u_{8}k+  u_{9}o - u_{10}v + u_{11}u ,\\
		k_{x}=2\lambda k - u_{1}a- u_{1}c-u_{5}g - u_{7}j- u_{9}d - u_{11}f,\\
		l_{x}=- 2\lambda l+u_{2}a + u_{2}c+ u_{6}f + u_{8}h + u_{10}e + u_{12}g ,\\
		m_{x}=2\lambda m- 2u_{3}b - 2u_{7}e - 2u_{9}h ,\\
		o_{x}=- 2\lambda o+2u_{4}b + 2u_{8}d + 2u_{10}j ,\\
		p_{x}=2\lambda p - 2u_{1}f - 2u_{5}a - 2u_{7}d,\\
		q_{x}=- 2\lambda q+ 2u_{2}g+2u_{6}a + 2u_{8}e  ,\\
		r_{x}=2\lambda r -u_{1}h- u_{3}d - u_{5}e- u_{7}a- u_{7}b  - u_{9}f ,\\
		s_{x}=- 2\lambda s+ u_{2}j+u_{4}e+ u_{6}d+u_{8}a + u_{8}b  + u_{10}g  ,\\
		t_{x}=2\lambda t-u_{1}e-u_{3}j-u_{7}g-u_{9}b- u_{9}c - u_{11}h ,\\
		u_{x}=- 2\lambda u+ u_{2}d+u_{4}h+u_{8}f +u_{10}b + u_{10}c + u_{12}j ,\\
		v_{x}=2\lambda v-2u_{1}g-2u_{9}j-2u_{11}c,\\
		w_{x}=- 2\lambda w+ 2u_{2}f+2u_{10}h+2u_{12}c  ,\\
	\end{cases}\label{eq:2.4}
\end{align}
Take the initial values
\begin{align*}
	a_{0}=\alpha, b_{0}=\beta, c_{0}=\gamma, d_{0}=e_{0}=\dots=v_{0}=w_{0}=0.
\end{align*}
From (\ref{eq:2.4}), the first few sets can be computed as follows
\begin{align*}
	a_{1}=&b_{1}=c_{1}=0,\ d_{1}=\frac{1}{2}\partial^{-1}(u_{1}u_{10}+u_{4}u_{7}+u_{5}u_{8})(\beta-\alpha),\
	e_{1}=\frac{1}{2}\partial^{-1}(u_{2}u_{9}+u_{3}u_{8}+u_{6}u_{7})(\alpha-\beta),\\
	f_{1}=&\frac{1}{2}\partial^{-1}(u_{1}u_{12}+u_{2}u_{5}+u_{7}u_{10})(\gamma-\alpha),\
	g_{1}=\frac{1}{2}\partial^{-1}(u_{1}u_{6}+u_{2}u_{11}+u_{8}u_{9})(\alpha-\gamma),\\
	h_{1}=&\frac{1}{2}\partial^{-1}(u_{2}u_{7}+u_{3}u_{10}+u_{9}u_{12})(\gamma-\beta),\
	j_{1}=\frac{1}{2}\partial^{-1}(u_{1}u_{8}+u_{4}u_{9}+u_{10}u_{11})(\beta-\gamma),\\
	k_{1}=&\frac{1}{2}u_{1}(\alpha+\gamma),\ l_{1}=\frac{1}{2}u_{2}(\alpha+\gamma),
	m_{1}=u_{3}\beta,\ o_{1}=u_{4}\beta,\ p_{1}=u_{5}\alpha,\ q_{1}=u_{6}\alpha,\\ r_{1}=&\frac{1}{2}u_{7}(\alpha+\beta),\ s_{1}=\frac{1}{2}u_{8}(\alpha+\beta),\
	t_{1}=\frac{1}{2}u_{9}(\beta+\gamma),\ u_{1}=\frac{1}{2}u_{10}(\beta+\gamma),\
	v_{1}=u_{11}\gamma,\ w_{1}=u_{12}\gamma,\\
	k_{2}=&\frac{1}{4}u_{1x}(\alpha+\gamma)+\frac{1}{4}u_{9}\partial^{-1}(u_{1}u_{10}+u_{4}u_{7}+u_{5}u_{8})(\beta-\alpha)+\frac{1}{4}u_{11}\partial^{-1}(u_{1}u_{12}+u_{2}u_{5}+u_{7}u_{10})(\gamma-\alpha)\\
	&+\frac{1}{4}u_{5}\partial^{-1}(u_{1}u_{6}+u_{2}u_{11}+u_{8}u_{9})(\alpha-\gamma)+\frac{1}{4}u_{7}\partial^{-1}(u_{1}u_{8}+u_{4}u_{9}+u_{10}u_{11})(\beta-\gamma),\\  
	l_{2}=&-\frac{1}{4}u_{2x}(\alpha+\gamma)+\frac{1}{4}u_{10}\partial^{-1}(u_{2}u_{9}+u_{3}u_{8}+u_{6}u_{7})(\alpha-\beta)+\frac{1}{4}u_{6}\partial^{-1}(u_{1}u_{12}+u_{2}u_{5}+u_{7}u_{10})(\gamma-\alpha)\\
	&+\frac{1}{4}u_{12}\partial^{-1}(u_{1}u_{6}+u_{2}u_{11}+u_{8}u_{9})(\alpha-\gamma)+\frac{1}{4}u_{8}\partial^{-1}(u_{2}u_{7}+u_{3}u_{10}+u_{9}u_{12})(\gamma-\beta),\\
	m_{2}=&\frac{1}{2}u_{3x}\beta+\frac{1}{2}u_{7}\partial^{-1}(u_{2}u_{9}+u_{3}u_{8}+u_{6}u_{7})(\alpha-\beta)+\frac{1}{2}u_{9}\partial^{-1}(u_{2}u_{7}+u_{3}u_{10}+u_{9}u_{12})(\gamma-\beta),\\
	o_{2}=&-\frac{1}{2}u_{4x}\beta+\frac{1}{2}u_{8}\partial^{-1}(u_{1}u_{10}+u_{4}u_{7}+u_{5}u_{8})(\beta-\alpha)+\frac{1}{2}u_{10}\partial^{-1}(u_{1}u_{8}+u_{4}u_{9}+u_{10}u_{11})(\beta-\gamma),\\
	p_{2}=&\frac{1}{2}u_{5x}\alpha+\frac{1}{2}u_{7}\partial^{-1}(u_{1}u_{10}+u_{4}u_{7}+u_{5}u_{8})(\beta-\alpha)+\frac{1}{2}u_{1}\partial^{-1}(u_{1}u_{12}+u_{2}u_{5}+u_{7}u_{10})(\gamma-\alpha),\\
	q_{2}=&-\frac{1}{2}u_{6x}\alpha+\frac{1}{2}u_{8}\partial^{-1}(u_{2}u_{9}+u_{3}u_{8}+u_{6}u_{7})(\alpha-\beta)+\frac{1}{2}u_{2}\partial^{-1}(u_{1}u_{6}+u_{2}u_{11}+u_{8}u_{9})(\alpha-\gamma),\\
	r_{2}=&\frac{1}{4}u_{7x}(\alpha+\beta)+\frac{1}{4}u_{3}\partial^{-1}(u_{1}u_{10}+u_{4}u_{7}+u_{5}u_{8})(\beta-\alpha)+\frac{1}{4}u_{5}\partial^{-1}(u_{2}u_{9}+u_{3}u_{8}+u_{6}u_{7})(\alpha-\beta)\\
	&+\frac{1}{4}u_{9}\partial^{-1}(u_{1}u_{12}+u_{2}u_{5}+u_{7}u_{10})(\gamma-\alpha)+\frac{1}{4}u_{1}\partial^{-1}(u_{2}u_{7}+u_{3}u_{10}+u_{9}u_{12})(\gamma-\beta),\\
	s_{2}=&-\frac{1}{4}u_{8x}(\alpha+\beta)+\frac{1}{4}u_{6}\partial^{-1}(u_{1}u_{10}+u_{4}u_{7}+u_{5}u_{8})(\beta-\alpha)+\frac{1}{4}u_{4}\partial^{-1}(u_{2}u_{9}+u_{3}u_{8}+u_{6}u_{7})(\alpha-\beta)\\
	&+\frac{1}{4}u_{10}\partial^{-1}(u_{1}u_{6}+u_{2}u_{11}+u_{8}u_{9})(\alpha-\gamma)+\frac{1}{4}u_{2}\partial^{-1}(u_{1}u_{8}+u_{4}u_{9}+u_{10}u_{11})(\beta-\gamma),\\
	t_{2}=&\frac{1}{4}u_{9x}(\beta+\gamma)+\frac{1}{4}u_{1}\partial^{-1}(u_{2}u_{9}+u_{3}u_{8}+u_{6}u_{7})(\alpha-\beta)+\frac{1}{4}u_{7}\partial^{-1}(u_{1}u_{6}+u_{2}u_{11}+u_{8}u_{9})(\alpha-\gamma)\\
	&+\frac{1}{4}u_{11}\partial^{-1}(u_{2}u_{7}+u_{3}u_{10}+u_{9}u_{12})(\gamma-\beta)+\frac{1}{4}u_{3}\partial^{-1}(u_{1}u_{8}+u_{4}u_{9}+u_{10}u_{11})(\beta-\gamma),\\
	u_{2}=&-\frac{1}{4}u_{10x}(\beta+\gamma)+\frac{1}{4}u_{2}\partial^{-1}(u_{1}u_{10}+u_{4}u_{7}+u_{5}u_{8})(\beta-\alpha)+\frac{1}{4}u_{8}\partial^{-1}(u_{1}u_{12}+u_{2}u_{5}+u_{7}u_{10})(\gamma-\alpha)\\
	&+\frac{1}{4}u_{4}\partial^{-1}(u_{2}u_{7}+u_{3}u_{10}+u_{9}u_{12})(\gamma-\beta)+\frac{1}{4}u_{12}\partial^{-1}(u_{1}u_{8}+u_{4}u_{9}+u_{10}u_{11})(\beta-\gamma),\\
	v_{2}=&\frac{1}{2}u_{11x}\gamma+\frac{1}{2}u_{1}\partial^{-1}(u_{1}u_{6}+u_{2}u_{11}+u_{8}u_{9})(\alpha-\gamma)+\frac{1}{2}u_{9}\partial^{-1}(u_{1}u_{8}+u_{4}u_{9}+u_{10}u_{11})(\beta-\gamma),\\
	w_{2}=&-\frac{1}{2}u_{12x}\gamma+\frac{1}{2}u_{2}\partial^{-1}(u_{1}u_{12}+u_{2}u_{5}+u_{7}u_{10})(\gamma-\alpha)+\frac{1}{2}u_{10}\partial^{-1}(u_{2}u_{7}+u_{3}u_{10}+u_{9}u_{12})(\gamma-\beta),\\
	\dots&\dots
\end{align*}
Now, taking
\begin{align*}
	V_{0}^{n}=\lambda^{n}V_{0}=\sum_{i\geq0}(a_{i},\dots,w_{i})^{t}\lambda^{n-i},\quad V_{0,+}^{n}=\sum_{0}^{n}(a_{i},\dots,w_{i})^{t}\lambda^{n-i},\quad V_{0,-}^{n}=V_{0}^{n}-V_{0,+}^{n},
\end{align*} then the zero curvature equation $U_{0,t}-V_{0,+x}^{n}+[U_{0},V_{0,+}^{n}]$ leads to the following Lax integrable hierarchy\\
\begin{align*}
	u_{t_{n}}=\left(
	\begin{array}{c}
		u_{1}\\
		u_{2}\\
		u_{3}\\
		u_{4}\\
		u_{5}\\
		u_{6}\\
		u_{7}\\
		u_{8}\\
		u_{9}\\
		u_{10}\\
		u_{11}\\
		u_{12}
	\end{array}\right)_{t_{n}}=
	\left(
	\begin{array}{c}
		2k_{n+1}\\
		-2l_{n+1}\\
		2m_{n+1}\\
		-2o_{n+1}\\
		2p_{n+1}\\
		-2q_{n+1}\\
		2r_{n+1}\\
		-2s_{n+1}\\
		2t_{n+1}\\
		-2u_{n+1}\\
		2v_{n+1}\\
		-2w_{n+1}\\
	\end{array}\right)
\end{align*}
\begin{align}
	=\left(\begin{array}{cccccccccccc}
		0&1&0&0&0&0&0&0&0&0&0&0\\
		-1&0&0&0&0&0&0&0&0&0&0&0\\
		0&0&0&2&0&0&0&0&0&0&0&0\\
		0&0&-2&0&0&0&0&0&0&0&0&0\\
		0&0&0&0&0&2&0&0&0&0&0&0\\
		0&0&0&0&-2&0&0&0&0&0&0&0\\
		0&0&0&0&0&0&0&1&0&0&0&0\\
		0&0&0&0&0&0&-1&0&0&0&0&0\\
		0&0&0&0&0&0&0&0&0&1&0&0\\
		0&0&0&0&0&0&0&0&-1&0&0&0\\
		0&0&0&0&0&0&0&0&0&0&0&2\\
		0&0&0&0&0&0&0&0&0&0&-2&0
	\end{array}\right)
	\left(
	\begin{array}{c}
		2l_{n+1}\\
		2k_{n+1}\\
		o_{n+1}\\
		m_{n+1}\\
		q_{n+1}\\
		p_{n+1}\\
		2s_{n+1}\\
		2r_{n+1}\\
		2u_{n+1}\\
		2t_{n+1}\\
		w_{n+1}\\
		v_{n+1}\\
	\end{array}\right)=
	J_{1}P_{1,n+1}.\label{eq:2.5}
\end{align}
From the recurrence relations (\ref{eq:2.4}), we have
	
	\begin{align*}
		P_{1,n+1}=(l_{i,j})_{12\times 12}P_{1,n}=L_{1}P_{1,n},
	\end{align*}
	where $L_{1}$ is a recurrence operator, and
	\begin{align*}
		&l_{1,1}=-\frac{1}{2}\partial+u_{2}\partial^{-1}u_{1}+\frac{1}{2}(u_{6}\partial^{-1}u_{5}+u_{8}\partial^{-1}u_{7}+u_{10}\partial^{-1}u_{9}+u_{12}\partial^{-1}u_{11}),\\ &l_{1,2}=-u_{2}\partial^{-1}u_{2}-\frac{1}{2}(u_{6}\partial^{-1}u_{12}+u_{12}\partial^{-1}u_{6}),\ l_{1,3}=0,\l_{1,4}=-(u_{8}\partial^{-1}u_{10}+u_{10}\partial^{-1}u_{8}),\\
		&l_{1,5}=u_{2}\partial^{-1}u_{5}+u_{10}\partial^{-1}u_{7}+u_{12}\partial^{-1}u_{1},\l_{1,6}=-(u_{2}\partial^{-1}u_{6}+u_{6}\partial^{-1}u_{2}),\\
		&l_{1,7}=\frac{1}{2}(u_{2}\partial^{-1}u_{7}+u_{10}\partial^{-1}u_{3}+u_{12}\partial^{-1}u_{9}),\ l_{1,8}=-\frac{1}{2}(u_{2}\partial^{-1}u_{8}+u_{6}\partial^{-1}u_{10}+u_{10}\partial^{-1}u_{6}+u_{8}\partial^{-1}u_{2}),\\
		&l_{1,9}=\frac{1}{2}(u_{2}\partial^{-1}u_{9}+u_{6}\partial^{-1}u_{7}+u_{8}\partial^{-1}u_{3}),\ 1_{1,10}=-\frac{1}{2}(u_{2}\partial^{-1}u_{10}+u_{8}\partial^{-1}u_{12}+u_{10}\partial^{-1}u_{2}+u_{12}\partial^{-1}u_{8}),\\
		&l_{1,11}=u_{2}\partial^{-1}u_{11}+u_{6}\partial^{-1}u_{1}+u_{8}\partial^{-1}u_{9},\ l_{1,12}=-(u_{2}\partial^{-1}u_{12}+u_{12}\partial^{-1}u_{2}),\\
		&l_{2,1}=u_{1}\partial^{-1}u_{1}+\frac{1}{2}(u_{5}\partial^{-1}u_{11}+u_{11}\partial^{-1}u_{5}),\\
		&l_{2,2}=\frac{1}{2}\partial-u_{1}\partial^{-1}u_{2}-\frac{1}{2}(u_{5}\partial^{-1}u_{6}+u_{7}\partial^{-1}u_{8}+u_{9}\partial^{-1}u_{10}+u_{11}\partial^{-1}u_{12}), \l_{2,3}=u_{7}\partial^{-1}u_{9}+u_{9}\partial^{-1}u_{7},\\
		&l_{2,4}=0,\ l_{2,5}=u_{1}\partial^{-1}u_{5}+u_{5}\partial^{-1}u_{1},\ l_{2,6}=-(u_{1}\partial^{-1}u_{6}+u_{9}\partial^{-1}u_{8}+u_{11}\partial^{-1}u_{2}),\\
		&l_{2,7}=\frac{1}{2}(u_{1}\partial^{-1}u_{7}+u_{5}\partial^{-1}u_{9}+u_{7}\partial^{-1}u_{1}+u_{9}\partial^{-1}u_{5}),\ l_{2,8}=-\frac{1}{2}(u_{1}\partial^{-1}u_{8}+u_{9}\partial^{-1}u_{4}+u_{11}\partial^{-1}u_{10}),\\
		&l_{2,9}=\frac{1}{2}(u_{1}\partial^{-1}u_{9}+u_{7}\partial^{-1}u_{11}+u_{9}\partial^{-1}u_{1}+u_{11}\partial^{-1}u_{7}),\ l_{2,10}=-\frac{1}{2}(u_{1}\partial^{-1}u_{10}+u_{5}\partial^{-1}u_{8}+u_{7}\partial^{-1}u_{4}),\\
		&l_{2,11}=u_{1}\partial^{-1}u_{11}+u_{11}\partial^{-1}u_{1},\ l_{2,12}=-(u_{1}\partial^{-1}u_{12}+u_{5}\partial^{-1}u_{2}+u_{7}\partial^{-1}u_{10}),\ l_{3,1}=0,\\ &l_{3,2}=-\frac{1}{2}(u_{8}\partial^{-1}u_{10}+u_{10}\partial^{-1}u_{8}),\ l_{3,3}=-\frac{1}{2}\partial+(u_{4}\partial^{-1}u_{3}+u_{8}\partial^{-1}u_{7}+u_{10}\partial^{-1}u_{9}),\ l_{3,4}=-u_{4}\partial^{-1}u_{4},\\
		&l_{3,5}=0,\ l_{3,6}=-u_{8}\partial^{-1}u_{8},\ l_{3,7}=\frac{1}{2}(u_{4}\partial^{-1}u_{7}+u_{8}\partial^{-1}u_{4}+u_{10}\partial^{-1}u_{1}),\ l_{3,8}=-\frac{1}{2}(u_{4}\partial^{-1}u_{8}+u_{8}\partial^{-1}u_{4}),\\ &l_{3,9}=\frac{1}{2}(u_{4}\partial^{-1}u_{9}+u_{8}\partial^{-1}u_{1}+u_{10}\partial^{-1}u_{11}),\l_{3,10}=-\frac{1}{2}(u_{4}\partial^{-1}u_{10}+u_{10}\partial^{-1}u_{4}),\ l_{3,11}=0,\\
		&l_{3,12}=-u_{10}\partial^{-1}u_{10},\ l_{4,1}=\frac{1}{2}(u_{7}\partial^{-1}u_{9}+u_{9}\partial^{-1}u_{7}),\ l_{4,2}=0,\ l_{4,3}=u_{3}\partial^{-1}u_{3},\\
		&l_{4,4}=\frac{1}{2}\partial-(u_{3}\partial^{-1}u_{4}+u_{7}\partial^{-1}u_{8}+u_{9}\partial^{-1}u_{10}),\ l_{4,5}=u_{7}\partial^{-1}u_{7},\ l_{4,6}=0,\ l_{4,7}=\frac{1}{2}(u_{3}\partial^{-1}u_{7}+u_{7}\partial^{-1}u_{3}),\\ &l_{4,8}=-\frac{1}{2}(u_{3}\partial^{-1}u_{8}+u_{7}\partial^{-1}u_{6}+u_{9}\partial^{-1}u_{2}),\ l_{4,9}=\frac{1}{2}(u_{3}\partial^{-1}u_{9}+u_{9}\partial^{-1}u_{3}),\\ &l_{4,10}=-\frac{1}{2}(u_{3}\partial^{-1}u_{10}+u_{7}\partial^{-1}u_{2}+u_{9}\partial^{-1}u_{12}),\ l_{4,11}=2u_{9}\partial^{-1}u_{9},\ l_{4,12}=0,\\
		&l_{5,1}=\frac{1}{2}(u_{6}\partial^{-1}u_{1}+u_{8}\partial^{-1}u_{9}+u_{2}\partial^{-1}u_{11}),\ l_{5,2}=-\frac{1}{2}(u_{6}\partial^{-1}u_{2}+u_{2}\partial^{-1}u_{6}),\ l_{5,3}=0,\ l_{5,4}=-u_{8}\partial^{-1}u_{8},\\ &l_{5,5}=-\frac{1}{2}\partial+u_{2}\partial^{-1}u_{1}+u_{6}\partial^{-1}u_{5}+u_{8}\partial^{-1}u_{7},\ l_{5,6}=-u_{6}\partial^{-1}u_{6},\ l_{5,7}=\frac{1}{2}(u_{2}\partial^{-1}u_{9}+u_{6}\partial^{-1}u_{7}+u_{8}\partial^{-1}u_{3}),\\
		&l_{5,8}=-\frac{1}{2}(u_{6}\partial^{-1}u_{8}+u_{8}\partial^{-1}u_{6}),\ l_{5,9}=0,\ l_{5,10}=-\frac{1}{2}(u_{2}\partial^{-1}u_{8}+u_{8}\partial^{-1}u_{2}),\ l_{5,11}=0,\ l_{5,12}=-u_{2}\partial^{-1}u_{2},\\
		&l_{6,1}=\frac{1}{2}(u_{1}\partial^{-1}u_{5}+u_{5}\partial^{-1}u_{1}),\ l_{6,2}=-\frac{1}{2}(u_{1}\partial^{-1}u_{12}+u_{5}\partial^{-1}u_{2}+u_{7}\partial^{-1}u_{10}),\ l_{6,3}=u_{7}\partial^{-1}u_{7},\ l_{6,4}=0,\\ &l_{6,5}=u_{5}\partial^{-1}u_{5},\ l_{6,6}=\frac{1}{2}\partial-u_{1}\partial^{-1}u_{2}+u_{5}\partial^{-1}u_{6}+u_{7}\partial^{-1}u_{8},\ l_{6,7}=\frac{1}{2}(u_{5}\partial^{-1}u_{7}+u_{7}\partial^{-1}u_{5}),\\ &l_{6,8}=-\frac{1}{2}(u_{1}\partial^{-1}u_{10}+u_{5}\partial^{-1}u_{8}+u_{7}\partial^{-1}u_{4}),\ l_{6,9}=\frac{1}{2}(u_{1}\partial^{-1}u_{7}+u_{7}\partial^{-1}u_{1}),\ l_{6,10}=0,\ l_{6,11}=u_{1}\partial^{-1}u_{1},\\
		&l_{6,12}=0,\ l_{7,1}=\frac{1}{2}(u_{4}\partial^{-1}u_{9}+u_{8}\partial^{-1}u_{1}+u_{10}\partial^{-1}u_{11}),\\ &l_{7,2}=-\frac{1}{2}(u_{2}\partial^{-1}u_{8}+u_{6}\partial^{-1}u_{10}+u_{8}\partial^{-1}u_{2}+u_{10}\partial^{-1}u_{6}),\ l_{7,3}=u_{2}\partial^{-1}u_{9}+u_{6}\partial^{-1}u_{7}+u_{8}\partial^{-1}u_{3},\\ &l_{7,4}=-(u_{4}\partial^{-1}u_{8}+u_{8}\partial^{-1}u_{4}),\ l_{7,5}=u_{4}\partial^{-1}u_{7}+u_{8}\partial^{-1}u_{5}+u_{10}\partial^{-1}u_{1},\ l_{7,6}=-(u_{6}\partial^{-1}u_{8}+u_{8}\partial^{-1}u_{6}),\\
		&l_{7,7}=-\frac{1}{2}\partial+u_{8}\partial^{-1}u_{7}+\frac{1}{2}(u_{2}\partial^{-1}u_{1}+u_{4}\partial^{-1}u_{3}+u_{6}\partial^{-1}u_{5}+u_{10}\partial^{-1}u_{9}),\\ &l_{7,8}=-u_{8}\partial^{-1}u_{8}-\frac{1}{2}(u_{4}\partial^{-1}u_{6}+u_{6}\partial^{-1}u_{4}),\ l_{7,9}=\frac{1}{2}(u_{2}\partial^{-1}u_{11}+u_{6}\partial^{-1}u_{1}+u_{8}\partial^{-1}u_{9}),\\ &l_{7,10}=-\frac{1}{2}(u_{2}\partial^{-1}u_{4}+u_{4}\partial^{-1}u_{2}+u_{8}\partial^{-1}u_{10}+u_{10}\partial^{-1}u_{8}),\ l_{7,11}=0,\ l_{7,12}=-(u_{2}\partial^{-1}u_{10}+u_{10}\partial^{-1}u_{2}),\\
		&l_{8,1}=\frac{1}{2}(u_{1}\partial^{-1}u_{7}+u_{5}\partial^{-1}u_{9}+u_{7}\partial^{-1}u_{1}+u_{9}\partial^{-1}u_{5}),\ l_{8,2}=-\frac{1}{2}(u_{3}\partial^{-1}u_{10}+u_{7}\partial^{-1}u_{2}+u_{9}\partial^{-1}u_{12}),\\
		&l_{8,3}=u_{3}\partial^{-1}u_{7}+u_{7}\partial^{-1}u_{3},\ l_{8,4}=-(u_{1}\partial^{-1}u_{10}+u_{5}\partial^{-1}u_{8}+u_{7}\partial^{-1}u_{4}),\ l_{8,5}=u_{5}\partial^{-1}u_{7}+u_{7}\partial^{-1}u_{5},\\ &l_{8,6}=-(u_{3}\partial^{-1}u_{8}+u_{7}\partial^{-1}u_{6}+u_{9}\partial^{-1}u_{2}),\ l_{8,7}=u_{7}\partial^{-1}u_{7}+\frac{1}{2}(u_{3}\partial^{-1}u_{5}+u_{5}\partial^{-1}u_{3}),\\
		&l_{8,8}=\frac{1}{2}\partial-u_{7}\partial^{-1}u_{8}-\frac{1}{2}(u_{1}\partial^{-1}u_{2}+u_{3}\partial^{-1}u_{4}+u_{5}\partial^{-1}u_{6}+u_{9}\partial^{-1}u_{10}),\\
		&l_{8,9}=\frac{1}{2}(u_{1}\partial^{-1}u_{3}+u_{3}\partial^{-1}u_{1}+u_{7}\partial^{-1}u_{9}+u_{9}\partial^{-1}u_{7}),\ l_{8,10}=-\frac{1}{2}(u_{1}\partial^{-1}u_{12}+u_{5}\partial^{-1}u_{2}+u_{7}\partial^{-1}u_{10}),\\
		&l_{8,11}=u_{1}\partial^{-1}u_{9}+u_{9}\partial^{-1}u_{1},\ l_{8,12}=0,\ l_{9,1}=\frac{1}{2}(u_{4}\partial^{-1}u_{7}+u_{8}\partial^{-1}u_{5}+u_{10}\partial^{-1}u_{1}),\\ &l_{9,2}=-\frac{1}{2}(u_{2}\partial^{-1}u_{10}+u_{8}\partial^{-1}u_{12}+u_{10}\partial^{-1}u_{2}+u_{12}\partial^{-1}u_{8}),\ l_{9,3}=u_{2}\partial^{-1}u_{7}+u_{10}\partial^{-1}u_{3}+u_{12}\partial^{-1}u_{9},\\ &l_{9,4}=-(u_{4}\partial^{-1}u_{10}+u_{10}\partial^{-1}u_{4}),\ l_{9,5}=0,\ l_{9,6}=-(u_{2}\partial^{-1}u_{8}+u_{8}\partial^{-1}u_{2}),\\ &l_{9,7}=\frac{1}{2}(u_{2}\partial^{-1}u_{5}+u_{10}\partial^{-1}u_{7}+u_{12}\partial^{-1}u_{1}),\ l_{9,8}=-\frac{1}{2}(u_{2}\partial^{-1}u_{4}+u_{4}\partial^{-1}u_{2}+u_{8}\partial^{-1}u_{10}+u_{10}\partial^{-1}u_{8}),\\
		&l_{9,9}=-\frac{1}{2}\partial+u_{10}\partial^{-1}u_{9}+\frac{1}{2}(u_{2}\partial^{-1}u_{1}+u_{4}\partial^{-1}u_{3}+u_{8}\partial^{-1}u_{7}+u_{12}\partial^{-1}u_{11}),\\
		&l_{9,10}=-u_{10}\partial^{-1}u_{10}-\frac{1}{2}(u_{4}\partial^{-1}u_{12}+u_{12}\partial^{-1}u_{4}),\ l_{9,11}=u_{4}\partial^{-1}u_{9}+u_{8}\partial^{-1}u_{1}+u_{10}\partial^{-1}u_{11},\\
		&l_{9,12}=-(u_{10}\partial^{-1}u_{12}+u_{12}\partial^{-1}u_{10}),\ l_{10,1}=\frac{1}{2}(u_{1}\partial^{-1}u_{9}+u_{7}\partial^{-1}u_{11}+u_{9}\partial^{-1}u_{1}+u_{11}\partial^{-1}u_{7}),\\
		&l_{10,2}=-\frac{1}{2}(u_{3}\partial^{-1}u_{8}+u_{7}\partial^{-1}u_{6}+u_{9}\partial^{-1}u_{2}),\ l_{10,3}=u_{3}\partial^{-1}u_{9}+u_{9}\partial^{-1}u_{3},\\
		&l_{10,4}=-(u_{1}\partial^{-1}u_{8}+u_{9}\partial^{-1}u_{4}+u_{11}\partial^{-1}u_{10}),\ l_{10,5}=u_{1}\partial^{-1}u_{7}+u_{7}\partial^{-1}u_{1},\ l_{10,6}=0,\\
		&l_{10,7}=\frac{1}{2}(u_{1}\partial^{-1}u_{3}+u_{3}\partial^{-1}u_{1}+u_{7}\partial^{-1}u_{9}+u_{9}\partial^{-1}u_{7}),\ l_{10,8}=-\frac{1}{2}(u_{1}\partial^{-1}u_{6}+u_{9}\partial^{-1}u_{8}+u_{11}\partial^{-1}u_{2}),\\
		&l_{10,9}=u_{9}\partial^{-1}u_{9}+\frac{1}{2}(u_{3}\partial^{-1}u_{11}+u_{11}\partial^{-1}u_{3}),\\
		&l_{10,10}=\frac{1}{2}\partial-u_{9}\partial^{-1}u_{10}-\frac{1}{2}(u_{1}\partial^{-1}u_{2}+u_{3}\partial^{-1}u_{4}+u_{7}\partial^{-1}u_{8}+u_{11}\partial^{-1}u_{12}),\\
		&l_{10,11}=u_{9}\partial^{-1}u_{11}+u_{11}\partial^{-1}u_{9},\ l_{10,12}=-(u_{3}\partial^{-1}u_{10}+u_{7}\partial^{-1}u_{2}+u_{9}\partial^{-1}u_{12}),\\
		&l_{11,1}=\frac{1}{2}(u_{2}\partial^{-1}u_{5}+u_{10}\partial^{-1}u_{7}+u_{12}\partial^{-1}u_{1}),\ l_{11,2}=-\frac{1}{2}(u_{2}\partial^{-1}u_{12}+u_{12}\partial^{-1}u_{2}),\ l_{11,3}=0,\\
		&l_{11,4}=-u_{10}\partial^{-1}u_{10},\ l_{11,5}=0,\ l_{11,6}=-u_{2}\partial^{-1}u_{2},\ l_{11,7}=0,\ l_{11,8}=-\frac{1}{2}(u_{2}\partial^{-1}u_{10}+u_{10}\partial^{-1}u_{2}),\\
		&l_{11,9}=\frac{1}{2}(u_{2}\partial^{-1}u_{7}+u_{10}\partial^{-1}u_{3}+u_{12}\partial^{-1}u_{9}),\ l_{11,10}=-\frac{1}{2}(u_{10}\partial^{-1}u_{12}+u_{12}\partial^{-1}u_{10}),\\
		&l_{11,11}=-\frac{1}{2}\partial+u_{2}\partial^{-1}u_{1}+u_{10}\partial^{-1}u_{9}+u_{12}\partial^{-1}u_{11},\ l_{11,12}=-u_{12}\partial^{-1}u_{12},\\ &l_{12,1}=\frac{1}{2}(u_{1}\partial^{-1}u_{11}+u_{11}\partial^{-1}u_{1}),\ l_{12,2}=-\frac{1}{2}(u_{1}\partial^{-1}u_{6}+u_{9}\partial^{-1}u_{8}+u_{11}\partial^{-1}u_{2}),\ l_{12,3}=u_{9}\partial^{-1}u_{9},\\
		&l_{12,4}=0,\ l_{12,5}=u_{1}\partial^{-1}u_{1},\ l_{12,6}=0,\ l_{12,7}=\frac{1}{2}(u_{1}\partial^{-1}u_{9}+u_{9}\partial^{-1}u_{1}),\ l_{12,8}=0,\\
		&l_{12,9}=\frac{1}{2}(u_{9}\partial^{-1}u_{11}+u_{11}\partial^{-1}u_{9}),\ l_{12,10}=-\frac{1}{2}(u_{1}\partial^{-1}u_{8}+u_{9}\partial^{-1}u_{4}+u_{11}\partial^{-1}u_{10}),\ l_{12,11}=u_{11}\partial^{-1}u_{11},\\
		&l_{12,12}=\frac{1}{2}\partial-(u_{1}\partial^{-1}u_{2}+u_{9}\partial^{-1}u_{10}+u_{11}\partial^{-1}u_{12}).
	\end{align*}
	We derive the Hamiltonian structure of (\ref{eq:2.5}) via the trace identity\cite{ref13}.
	\begin{align*}
		&\left\langle V_{0},\frac{\partial U_{0}}{\partial\lambda}\right\rangle = 2a+2b+2c ,\
		\left\langle V_{0},\frac{\partial U_{0}}{\partial u_{1}}\right\rangle = 2l ,\
		\left\langle V_{0},\frac{\partial U_{0}}{\partial u_{2}}\right\rangle = 2k ,\
		\left\langle V_{0},\frac{\partial U_{0}}{\partial u_{3}}\right\rangle = o ,\\
		&\left\langle V_{0},\frac{\partial U_{0}}{\partial u_{4}}\right\rangle = m ,\
		\left\langle V_{0},\frac{\partial U_{0}}{\partial u_{5}}\right\rangle = q ,\
		\left\langle V_{0},\frac{\partial U_{0}}{\partial u_{6}}\right\rangle = p ,\
		\left\langle V_{0},\frac{\partial U_{0}}{\partial u_{7}}\right\rangle = 2s ,\
		\left\langle V_{0},\frac{\partial U_{0}}{\partial u_{8}}\right\rangle = 2r ,\\
		&\left\langle V_{0},\frac{\partial U_{0}}{\partial u_{9}}\right\rangle = 2u ,\
		\left\langle V_{0},\frac{\partial U_{0}}{\partial u_{10}}\right\rangle = 2t ,\
		\left\langle V_{0},\frac{\partial U_{0}}{\partial u_{11}}\right\rangle = w ,\
		\left\langle V_{0},\frac{\partial U_{0}}{\partial u_{12}}\right\rangle = v ,\
	\end{align*}
	Substituting the above formulate into the trace identity yields
	\begin{align*}
		\frac{\delta}{\delta u}(2a+2b+2c)=\lambda^{-\tau}\frac{\partial}{\partial\lambda}\lambda^{\tau}
		\left(\begin{array}{c}
			2l\\
			2k\\
			o\\
			m\\
			q\\
			p\\
			2s\\
			2r\\
			2u\\
			2t\\
			w\\
			v
		\end{array}\right),
	\end{align*}
	where $\tau=\frac{\lambda}{2}\frac{d}{dx}ln|tr(V_{0}^{2})|$. Balancing coefficients of each power of  in the above equality gives rise to
	\begin{align*}
		\frac{\delta}{\delta u}(2a_{n+1}+2b_{n+1}+2c_{n+1})=(\tau-n)
		\left(\begin{array}{c}
			2l_{n}\\
			2k_{n}\\
			o_{n}\\
			m_{n}\\
			q_{n}\\
			p_{n}\\
			2s_{n}\\
			2r_{n}\\
			2u_{n}\\
			2t_{n}\\
			w_{n}\\
			v_{n}
		\end{array}\right),
	\end{align*}
	Taking $n=1$, gives $\tau=-1$. Therefore we establish the following equation:
	\begin{align*}
		P_{1,n+1}=
		\left(\begin{array}{c}
			2l_{n+1}\\
			2k_{n+1}\\
			o_{n+1}\\
			m_{n+1}\\
			q_{n+1}\\
			p_{n+1}\\
			2s_{n+1}\\
			2r_{n+1}\\
			2u_{n+1}\\
			2t_{n+1}\\
			w_{n+1}\\
			v_{n+1}
		\end{array}\right)=\frac{\delta}{\delta u}((\frac{-2}{n+2})(a_{n+2}+b_{n+2}+c_{n+2})).
	\end{align*}
	Thus, we see:
	\begin{align*}
		u_{t}=J_{1}P_{1,n+1}=J_{1}\frac{\delta H_{n+1}^{1}}{\delta u},\  H_{n+1}^{1}=(\frac{-2}{n+2})(a_{n+2}+b_{n+2}+c_{n+2}),\ n\geq0.
	\end{align*}
	It is said that the hierarchy (\ref{eq:2.5}) have the Hamiltonian structure, and it is easy to verity that $J_{1}L_{1}=L_{1}^{*}J_{1}$. Therefore, the hierarchy (\ref{eq:2.5}) is Liouville integrable.

	When $n=1$, the hierarchy (\ref{eq:2.5}) reduces to the first integrable system
	\begin{align}\label{eq:2.6}
		u_{1t}=&\frac{1}{2}u_{1x}(\alpha+\gamma)+\frac{1}{2}u_{9}\partial^{-1}(u_{1}u_{10}+u_{4}u_{7}+u_{5}u_{8})(\beta-\alpha)+\frac{1}{2}u_{11}\partial^{-1}(u_{1}u_{12}+u_{2}u_{5}+u_{7}u_{10})(\gamma-\alpha)\notag\\
		&+\frac{1}{2}u_{5}\partial^{-1}(u_{1}u_{6}+u_{2}u_{11}+u_{8}u_{9})(\alpha-\gamma)+\frac{1}{2}u_{7}\partial^{-1}(u_{1}u_{8}+u_{4}u_{9}+u_{10}u_{11})(\beta-\gamma),\notag\\
		u_{2t}=&\frac{1}{2}u_{2x}(\alpha+\gamma)-\frac{1}{2}u_{10}\partial^{-1}(u_{2}u_{9}+u_{3}u_{8}+u_{6}u_{7})(\alpha-\beta)-\frac{1}{2}u_{6}\partial^{-1}(u_{1}u_{12}+u_{2}u_{5}+u_{7}u_{10})(\gamma-\alpha)\notag\\
		&-\frac{1}{2}u_{12}\partial^{-1}(u_{1}u_{6}+u_{2}u_{11}+u_{8}u_{9})(\alpha-\gamma)-\frac{1}{2}u_{8}\partial^{-1}(u_{2}u_{7}+u_{3}u_{10}+u_{9}u_{12})(\gamma-\beta),\notag\\
		u_{3t}=&u_{3x}\beta+u_{7}\partial^{-1}(u_{2}u_{9}+u_{3}u_{8}+u_{6}u_{7})(\alpha-\beta)+u_{9}\partial^{-1}(u_{2}u_{7}+u_{3}u_{10}+u_{9}u_{12})(\gamma-\beta),\notag\\
		u_{4t}=&u_{4x}\beta-u_{8}\partial^{-1}(u_{1}u_{10}+u_{4}u_{7}+u_{5}u_{8})(\beta-\alpha)-u_{10}\partial^{-1}(u_{1}u_{8}+u_{4}u_{9}+u_{10}u_{11})(\beta-\gamma),\notag\\
		u_{5t}=&u_{5x}\alpha+u_{7}\partial^{-1}(u_{1}u_{10}+u_{4}u_{7}+u_{5}u_{8})(\beta-\alpha)+u_{1}\partial^{-1}(u_{1}u_{12}+u_{2}u_{5}+u_{7}u_{10})(\gamma-\alpha),\notag\\
		u_{6t}=&u_{6x}\alpha-u_{8}\partial^{-1}(u_{2}u_{9}+u_{3}u_{8}+u_{6}u_{7})(\alpha-\beta)-u_{2}\partial^{-1}(u_{1}u_{6}+u_{2}u_{11}+u_{8}u_{9})(\alpha-\gamma),\notag\\
		u_{7t}=&\frac{1}{2}u_{7x}(\alpha+\beta)+\frac{1}{2}u_{3}\partial^{-1}(u_{1}u_{10}+u_{4}u_{7}+u_{5}u_{8})(\beta-\alpha)+\frac{1}{2}u_{5}\partial^{-1}(u_{2}u_{9}+u_{3}u_{8}+u_{6}u_{7})(\alpha-\beta)\notag\\
		&+\frac{1}{2}u_{9}\partial^{-1}(u_{1}u_{12}+u_{2}u_{5}+u_{7}u_{10})(\gamma-\alpha)+\frac{1}{2}u_{1}\partial^{-1}(u_{2}u_{7}+u_{3}u_{10}+u_{9}u_{12})(\gamma-\beta),\notag\\
		u_{8t}=&\frac{1}{2}u_{8x}(\alpha+\beta)-\frac{1}{2}u_{6}\partial^{-1}(u_{1}u_{10}+u_{4}u_{7}+u_{5}u_{8})(\beta-\alpha)-\frac{1}{2}u_{4}\partial^{-1}(u_{2}u_{9}+u_{3}u_{8}+u_{6}u_{7})(\alpha-\beta)\\
		&-\frac{1}{2}u_{10}\partial^{-1}(u_{1}u_{6}+u_{2}u_{11}+u_{8}u_{9})(\alpha-\gamma)-\frac{1}{2}u_{2}\partial^{-1}(u_{1}u_{8}+u_{4}u_{9}+u_{10}u_{11})(\beta-\gamma),\notag\\
		u_{9t}=&\frac{1}{2}u_{9x}(\beta+\gamma)+\frac{1}{2}u_{1}\partial^{-1}(u_{2}u_{9}+u_{3}u_{8}+u_{6}u_{7})(\alpha-\beta)+\frac{1}{2}u_{7}\partial^{-1}(u_{1}u_{6}+u_{2}u_{11}+u_{8}u_{9})(\alpha-\gamma)\notag\\
		&+\frac{1}{2}u_{11}\partial^{-1}(u_{2}u_{7}+u_{3}u_{10}+u_{9}u_{12})(\gamma-\beta)+\frac{1}{2}u_{3}\partial^{-1}(u_{1}u_{8}+u_{4}u_{9}+u_{10}u_{11})(\beta-\gamma),\notag\\
		u_{10t}=&\frac{1}{2}u_{10x}(\beta+\gamma)-\frac{1}{2}u_{2}\partial^{-1}(u_{1}u_{10}+u_{4}u_{7}+u_{5}u_{8})(\beta-\alpha)-\frac{1}{2}u_[8]\partial^{-1}(u_{1}u_{12}+u_{2}u_{5}+u_{7}u_{10})(\gamma-\alpha)\notag\\
		&-\frac{1}{2}u_{4}\partial^{-1}(u_{2}u_{7}+u_{3}u_{10}+u_{9}u_{12})(\gamma-\beta)-\frac{1}{2}u_{12}\partial^{-1}(u_{1}u_{8}+u_{4}u_{9}+u_{10}u_{11})(\beta-\gamma),\notag\\
		u_{11t}=&u_{11x}\gamma+u_{1}\partial^{-1}(u_{1}u_{6}+u_{2}u_{11}+u_{8}u_{9})(\alpha-\gamma)+u_{9}\partial^{-1}(u_{1}u_{8}+u_{4}u_{9}+u_{10}u_{11})(\beta-\gamma),\notag\\
		u_{12t}=&u_{12x}\gamma-u_{2}\partial^{-1}(u_{1}u_{12}+u_{2}u_{5}+u_{7}u_{10})(\gamma-\alpha)-u_{10}\partial^{-1}(u_{2}u_{7}+u_{3}u_{10}+u_{9}u_{12})(\gamma-\beta)\notag
	\end{align}
	At this stage, we have successfully constructed an integrable soliton hierarchy and, as a concrete illustration, presented its integrable system for the case when $n=1$.

	\subsection{ The Second Soliton Hierarchy Associated with $\mathfrak{sp}(6)$.\\}\label{Sec:2-2}
	In this section, we are going to construct another soliton hierarchy from the loop algebra $\widetilde{\mathfrak{sp}(6)}$. We consider another isospectral problem
	\begin{align*}
		\begin{cases}
			\phi_{x} = U_{1}\phi, \\
			\phi_{t} = V_{0}\phi, \quad \lambda_{t} = 0.
		\end{cases}
	\end{align*}
	
	Let $U_{1}\in \widetilde{\mathfrak{sp}(6)}$ be given as follows
	\begin{align*}
		U_{1}=&E_{1}(1)-E_{2}(1)-E_{3}(1)+u_{1}'E_{6}(0)+u_{2}'E_{7}(0)+u_{3}'E_{12}(0)+u_{4}'E_{13}(0)+u_{5}'E_{14}(0)\\
		&+u_{6}'E_{15}(0)+u_{7}'E_{4}(0)+u_{8}'E_{5}(0)+u_{9}'E_{18}(0)+u_{10}'E_{19}(0)+u_{11}'E_{20}(0)+u_{12}'E_{21}(0),
	\end{align*}
	i.e.
	\begin{align}
		U_{1}=\left(
		\begin{array}{cccccc}
			\lambda&u_{7}'&u_{1}'&u_{5}'&0&0\\
			u_{8}'&-\lambda&0&0&u_{3}'&u_{9}'\\
			u_{2}'&0&-\lambda&0&u_{9}'&u_{11}'\\
			u_{6}'&0&0&-\lambda&-u_{8}'&-u_{2}'\\
			0&u_{4}'&u_{10}'&-u_{7}'&\lambda&0\\
			0&u_{10}'&u_{12}'&-u_{1}'&0&\lambda
		\end{array}\right)\label{eq:2.7}
	\end{align}
	The stationary zero curvature representation $V_{0,x}=[U_{1},V_{0}]$ gives
	\begin{align}
		\begin{cases}
			a_{x}=u_{1}'g-u_{2}'f+u_{5}'q-u_{6}'p+u_{7}'e-u_{8}'d,\\
			b_{x}=u_{3}'o-u_{4}'m-u_{7}'e+u_{8}'d+u_{9}'u-u_{10}'t,\\
			c_{x}=-u_{1}'g+u_{2}'f+u_{9}'u-u_{10}'t+u_{11}'w-u_{12}'v,\\
			d_{x}=2\lambda d+u_{1}'j-u_{4}'r+u_{5}'s-u_{7}'a+u_{7}'b-u_{10}'k,\\
			e_{x}=-2\lambda e-u_{2}'h+u_{3}'s-u_{6}'r+u_{8}'a-u_{8}'b+u_{9}'l,\\
			f_{x}=2\lambda f-u_{1}'a+u_{1}'c+u_{5}'l+u_{7}'h-u_{10}'r-u_{12}'k,\\
			g_{x}=-2\lambda g+u_{2}'a-u_{2}'c-u_{6}'k-u_{8}'j+u_{9}'s+u_{11}'l,\\
			h_{x}=-u_{1}'e+u_{3}'u+u_{8}'f+u_{9}'w-u_{10}'m-u_{12}'t,\\
			j_{x}=u_{2}'d-u_{4}'t-u_{7}'g+u_{9}'o-u_{10}'v+u_{11}'u,\\
			k_{x}=u_{1}'v+u_{2}'p-u_{5}'g+u_{7}'t-u_{9}'d-u_{11}'f,\\
			l_{x}=-u_{1}'q-u_{2}'w+u_{6}'f-u_{8}'u+u_{10}'e+u_{12}'g,\\
			m_{x}=-2\lambda m-2u_{3}'b+2u_{8}'r-2u_{9}'h,\\
			o_{x}=2\lambda o+2u_{4}'b-2u_{7}'s+2u_{10}'j,\\
			p_{x}=2\lambda p+2u_{1}'k-2u_{5}'a+2u_{7}'r,\\
			q_{x}=-2\lambda q-2u_{2}'l+2u_{6}'a-2u_{8}'s,\\
			r_{x}=u_{1}'t-u_{3}'d-u_{5}'e+u_{7}'m+u_{8}'p-u_{9}'f,\\
			s_{x}=-u_{2}'u+u_{4}'e+u_{6}'d-u_{7}'q-u_{8}'o+u_{10}'g,\\
			t_{x}=-2\lambda t+u_{2}'r-u_{3}'j+u_{8}'k-u_{9}'b-u_{9}'c-u_{11}'h,\\
			u_{x}=2\lambda u-u_{1}'s+u_{4}'h-u_{7}'l+u_{10}'b+u_{10}'c+u_{12}'j,\\
			v_{x}=-2\lambda v+2u_{2}'k-2u_{9}'j-2u_{11}'c,\\
			w_{x}=2\lambda w-2u_{1}'l+2u_{10}'h+2u_{12}'c.\\
		\end{cases}\label{eq:2.8}
	\end{align}
	Take the initial values
	\begin{align*}
		a_{0}=\alpha, b_{0}=\beta, c_{0}=\gamma, d_{0}=e_{0}=\dots=v_{0}=w_{0}=0.
	\end{align*}
	From (\ref{eq:2.7}), the first few sets can be computed as follows
	\begin{align*}
		a_{1}=&b_{1}=c_{1}=0,\ d_{1}=\frac{1}{2}u_{7}'(\alpha-\beta),\ e_{1}=\frac{1}{2}u_{8}'(\alpha-\beta),\ f_{1}=\frac{1}{2}u_{1}'(\alpha-\gamma),\ g_{1}=\frac{1}{2}u_{2}'(\alpha-\gamma),\\ h_{1}=&\frac{1}{2}\partial^{-1}(u_{1}'u_{8}'+u_{3}'u_{10}'+u_{9}'u_{12}')(\beta-\gamma),\ j_{1}=-\frac{1}{2}\partial^{-1}(u_{2}'u_{7}'+u_{4}'u_{9}'+u_{10}'u_{11}')(\beta-\gamma),\\ k_{1}=&\frac{1}{2}\partial^{-1}(-u_{1}'u_{11}'+u_{2}'u_{5}'-u_{7}'u_{9}')(\alpha+\gamma),\ l_{1}=\frac{1}{2}\partial^{-1}(-u_{1}'u_{6}'+u_{2}'u_{12}'+u_{8}'u_{10}')(\alpha+\gamma),\ m_{1}=-u_{3}'\beta,\\
		o_{1}=&-u_{4}'\beta,\ p_{1}=u_{5}'\alpha,\ q_{1}=u_{6}'\alpha,\ r_{1}=\frac{1}{2}\partial^{-1}(-u_{1}'u_{9}'-u_{3}'u_{7}'+u_{5}'u_{8}')(\alpha+\beta),\\ s_{1}=&\frac{1}{2}\partial^{-1}(u_{2}'u_{10}'+u_{4}'u_{8}'-u_{6}'u_{7}')(\alpha+\beta),\ t_{1}=-\frac{1}{2}u_{9}'(\beta+\gamma),\ u_{1}=-\frac{1}{2}u_{10}'(\beta+\gamma),\\ v_{1}=&-u_{11}'\gamma,\ w_{1}=-u_{12}'\gamma,\\ d_{2}=&\frac{1}{4}u_{7x}'(\alpha-\beta)+\frac{1}{4}u_{1}'\partial^{-1}(u_{2}'u_{7}'+u_{4}'u_{9}'+u_{10}'u_{11}')(\beta-\gamma)+\frac{1}{4}u_{10}'\partial^{-1}(-u_{1}'u_{11}'+u_{2}'u_{5}'-u_{7}'u_{9}')(\alpha+\gamma)\\
		&+\frac{1}{4}u_{4}'\partial^{-1}(-u_{1}'u_{9}'-u_{3}'u_{7}'+u_{5}'u_{8}')(\alpha+\beta)-\frac{1}{4}u_{5}'\partial^{-1}(u_{2}'u_{10}'+u_{4}'u_{8}'-u_{6}'u_{7}')(\alpha+\beta),\\
		e_{2}=&-\frac{1}{4}u_{8x}'(\alpha-\beta)-\frac{1}{4}u_{2}'\partial^{-1}(u_{1}'u_{8}'+u_{3}'u_{10}'+u_{9}'u_{12}')(\beta-\gamma)+\frac{1}{4}u_{9}'\partial^{-1}(-u_{1}'u_{6}'+u_{2}'u_{12}'+u_{8}'u_{10}')(\alpha+\gamma)\\
		&-\frac{1}{6}u_{4}'\partial^{-1}(-u_{1}'u_{9}'-u_{3}'u_{7}'+u_{5}'u_{8}')(\alpha+\beta)+\frac{1}{4}u_{3}'\partial^{-1}(u_{2}'u_{10}'+u_{4}'u_{8}'-u_{6}'u_{7}')(\alpha+\beta),\\
		f_{2}=&\frac{1}{4}u_{1x}'(\alpha-\gamma)-\frac{1}{4}u_{7}'\partial^{-1}(u_{1}'u_{8}'+u_{3}'u_{10}'+u_{9}'u_{12}')(\beta-\gamma)+\frac{1}{4}u_{12}'\partial^{-1}(-u_{1}'u_{11}'+u_{2}'u_{5}'-u_{7}'u_{9}')(\alpha+\gamma)\\
		&-\frac{1}{4}u_{5}'\partial^{-1}(-u_{1}'u_{6}'+u_{2}'u_{12}'+u_{8}'u_{10}')(\alpha+\gamma)+\frac{1}{4}u_{10}'\partial^{-1}(-u_{1}'u_{9}'-u_{3}'u_{7}'+u_{5}'u_{8}')(\alpha+\beta),\\
		g_{2}=&-\frac{1}{4}u_{2x}'(\alpha-\gamma)+\frac{1}{4}u_{8}'\partial^{-1}(u_{2}'u_{7}'+u_{4}'u_{9}'+u_{10}'u_{11}')(\beta-\gamma)-\frac{1}{4}u_{6}'\partial^{-1}(-u_{1}'u_{11}'+u_{2}'u_{5}'-u_{7}'u_{9}')(\alpha+\gamma)\\
		&+\frac{1}{4}u_{11}'\partial^{-1}(-u_{1}'u_{6}'+u_{2}'u_{12}'+u_{8}'u_{10}')(\alpha+\gamma)+\frac{1}{4}u_{9}'\partial^{-1}(u_{2}'u_{10}'+u_{4}'u_{8}'-u_{6}'u_{7}')(\alpha+\beta),\\
		m_{2}=&\frac{1}{2}u_{3x}'\beta+\frac{1}{2}u_{8}'\partial^{-1}(-u_{1}'u_{9}'-u_{3}'u_{7}'+u_{5}'u_{8}')(\alpha+\beta)-\frac{1}{2}u_{9}'\partial^{-1}(u_{1}'u_{8}'+u_{3}'u_{10}'+u_{9}'u_{12}')(\beta-\gamma),\\
		o_{2}=&-\frac{1}{2}u_{4x}'\beta+\frac{1}{2}u_{10}'\partial^{-1}(u_{2}'u_{7}'+u_{4}'u_{9}'+u_{10}'u_{11}')(\beta-\gamma)+\frac{1}{2}u_{7}'\partial^{-1}(u_{2}'u_{10}'+u_{4}'u_{8}'-u_{6}'u_{7}')(\alpha+\beta),\\
		p_{2}=&\frac{1}{2}u_{5x}'\alpha-\frac{1}{2}u_{1}'\partial^{-1}(-u_{1}'u_{11}'+u_{2}'u_{5}'-u_{7}'u_{9}')(\alpha+\gamma)-\frac{1}{2}u_{7}'\partial^{-1}(-u_{1}'u_{9}'-u_{3}'u_{7}'+u_{5}'u_{8}')(\alpha+\beta),\\
		q_{2}=&-\frac{1}{2}u_{6x}'\alpha-\frac{1}{2}u_{2}'\partial^{-1}(-u_{1}'u_{6}'+u_{2}'u_{12}'+u_{8}'u_{10}')(\alpha+\gamma)-\frac{1}{2}u_{8}'\partial^{-1}(u_{2}'u_{10}'+u_{4}'u_{8}'-u_{6}'u_{7}')(\alpha+\beta),\\
		t_{2}=&\frac{1}{4}u_{9x}'(\beta+\gamma)+\frac{1}{4}u_{8}'\partial^{-1}(-u_{1}'u_{11}'+u_{2}'u_{5}'-u_{7}'u_{9}')(\alpha+\gamma)-\frac{1}{4}u_{11}'\partial^{-1}(u_{1}'u_{8}'+u_{3}'u_{10}'+u_{9}'u_{12}')(\beta-\gamma),\\
		&+\frac{1}{4}u_{3}'\partial^{-1}(u_{2}'u_{7}'+u_{4}'u_{9}'+u_{10}'u_{11}')(\beta-\gamma)+\frac{1}{4}u_{2}'\partial^{-1}(-u_{1}'u_{9}'-u_{3}'u_{7}'+u_{5}'u_{8}')(\alpha+\beta),\\
		u_{2}=&-\frac{1}{4}u_{10x}'(\beta+\gamma)-\frac{1}{4}u_{4}'\partial^{-1}(u_{1}'u_{8}'+u_{3}'u_{10}'+u_{9}'u_{12}')(\beta-\gamma)+\frac{1}{4}u_{12}'\partial^{-1}(u_{2}'u_{7}'+u_{4}'u_{9}'+u_{10}'u_{11}')(\beta-\gamma),\\
		&+\frac{1}{4}u_{7}'\partial^{-1}(-u_{1}'u_{6}'+u_{2}'u_{12}'+u_{8}'u_{10}')(\alpha+\gamma)+\frac{1}{4}u_{1}'\partial^{-1}(u_{2}'u_{10}'+u_{4}'u_{8}'-u_{6}'u_{7}')(\alpha+\beta),\\
		v_{2}=&\frac{1}{2}u_{11x}'\gamma+\frac{1}{2}u_{2}'\partial^{-1}(-u_{1}'u_{11}'+u_{2}'u_{5}'-u_{7}'u_{9}')(\alpha+\gamma)+\frac{1}{2}u_{9}'\partial^{-1}(u_{2}'u_{7}'+u_{4}'u_{9}'+u_{10}'u_{11}')(\beta-\gamma),\\
		w_{2}=&-\frac{1}{2}u_{12x}'\gamma-\frac{1}{2}u_{10}'\partial^{-1}(u_{1}'u_{8}'+u_{3}'u_{10}'+u_{9}'u_{12}')(\beta-\gamma)+\frac{1}{2}u_{1}'\partial^{-1}(-u_{1}'u_{6}'+u_{2}'u_{12}'+u_{8}'u_{10}')(\alpha+\gamma),\\
		\dots&\dots
	\end{align*}
	The zero curvature equation $U_{1,t}-V_{0,+x}^{n}+[U_{1},V_{0,+}^{n}]$ leads to the following Lax integrable hierarchy
	\begin{align*}
		u_{t_{n}}'=\left(
		\begin{array}{c}
			u_{1}'\\
			u_{2}'\\
			u_{3}'\\
			u_{4}'\\
			u_{5}'\\
			u_{6}'\\
			u_{7}'\\
			u_{8}'\\
			u_{9}'\\
			u_{10}'\\
			u_{11}'\\
			u_{12}'
		\end{array}\right)_{t_{n}}=
		\left(\begin{array}{c}
			2f_{n+1}\\
			-2g_{n+1}\\
			-2m_{n+1}\\
			2o_{n+1}\\
			2p_{n+1}\\
			-2q_{n+1}\\
			2d_{n+1}\\
			-2e_{n+1}\\
			-2t_{n+1}\\
			2u_{n+1}\\
			-2v_{n+1}\\
			2w_{n+1}
		\end{array}\right)
	\end{align*}
	\begin{align}
		=\left(\begin{array}{cccccccccccc}
			0&1&0&0&0&0&0&0&0&0&0&0\\
			-1&0&0&0&0&0&0&0&0&0&0&0\\
			0&0&0&-2&0&0&0&0&0&0&0&0\\
			0&0&2&0&0&0&0&0&0&0&0&0\\
			0&0&0&0&0&2&0&0&0&0&0&0\\
			0&0&0&0&-2&0&0&0&0&0&0&0\\
			0&0&0&0&0&0&0&1&0&0&0&0\\
			0&0&0&0&0&0&-1&0&0&0&0&0\\
			0&0&0&0&0&0&0&0&0&-1&0&0\\
			0&0&0&0&0&0&0&0&1&0&0&0\\
			0&0&0&0&0&0&0&0&0&0&0&-2\\
			0&0&0&0&0&0&0&0&0&0&2&0
		\end{array}\right)
		\left(
		\begin{array}{c}
			2g_{n+1}\\
			2f_{n+1}\\
			o_{n+1}\\
			m_{n+1}\\
			q_{n+1}\\
			p_{n+1}\\
			2e_{n+1}\\
			2d_{n+1}\\
			2u_{n+1}\\
			2t_{n+1}\\
			w_{n+1}\\
			v_{n+1}\\
		\end{array}\right)=
		J_{2}P_{2,n+1}.\label{eq:2.9}
	\end{align}
	From the recurrence relations (\ref{eq:2.7}), we have
	\begin{align*}
		P_{2,n+1}=(l_{i,j}')_{12\times 12}P_{2,n}=L_{2}P_{2,n},
	\end{align*}
	where $L_{2}$ is a recurrence operator, and
	
	\begin{align*}
		&l_{1,1}'=-\frac{1}{2}\partial+u_{2}'\partial^{-1}u_{1}'+\frac{1}{2}(u_{6}'\partial^{-1}u_{5}'+u_{8}'\partial^{-1}u_{7}'+u_{9}'\partial^{-1}u_{10}'+u_{11}'\partial^{-1}u_{12}'),\\&
		l_{1,2}'=-u_{2}'\partial^{-1}u_{2}'+\frac{1}{2}(u_{6}'\partial^{-1}u_{11}'+u_{11}'\partial^{-1}u_{6}'),\ l_{1,3}'=-(u_{8}'\partial^{-1}u_{9}'+u_{9}'\partial^{-1}u_{8}'),\ l_{1,4}'=0,\\& l_{1,5}'=u_{2}'\partial^{-1}u_{5}'-u_{9}'\partial^{-1}u_{7}'-u_{11}'\partial^{-1}u_{1}',\ l_{1,6}'=-(u_{2}'\partial^{-1}u_{6}'+u_{6}'\partial^{-1}u_{2}'),\\& l_{1,7}'=\frac{1}{2}(u_{2}'\partial^{-1}u_{7}'+u_{9}'\partial^{-1}u_{4}'+u_{11}'\partial^{-1}u_{10}'),\ l_{1,8}'=\frac{1}{2}(-u_{2}'\partial^{-1}u_{8}'+u_{6}'\partial^{-1}u_{9}'-u_{8}'\partial^{-1}u_{2}'+u_{9}'\partial^{-1}u_{6}'),\\& l_{1,9}'=-\frac{1}{2}(u_{2}'\partial^{-1}u_{9}'+u_{8}'\partial^{-1}u_{11}'+u_{9}'\partial^{-1}u_{2}'+u_{11}'\partial^{-1}u_{8}'),\ l_{1,10}'=\frac{1}{2}(u_{2}'\partial^{-1}u_{10}'-u_{6}'\partial^{-1}u_{7}'+u_{8}'\partial^{-1}u_{4}'),\\& l_{1,11}'=-(u_{2}'\partial^{-1}u_{11}'+u_{11}'\partial^{-1}u_{2}'),\ l_{1,12}'=u_{2}'\partial^{-1}u_{12}'-u_{6}'\partial^{-1}u_{1}'+u_{8}'\partial^{-1}u_{12}',\\&
		l_{2,1}'=u_{1}'\partial^{-1}u_{1}'-\frac{1}{2}(u_{5}'\partial^{-1}u_{12}'+u_{12}'\partial^{-1}u_{5}'),\\& l_{2,2}'=\frac{1}{2}\partial-u_{1}'\partial^{-1}u_{2}'-\frac{1}{2}(u_{5}'\partial^{-1}u_{6}'+u_{7}'\partial^{-1}u_{8}'+u_{10}'\partial^{-1}u_{9}'+u_{12}'\partial^{-1}u_{11}'),\ l_{2,3}'=0,\\& l_{2,4}'=u_{7}'\partial^{-1}u_{10}'+u_{10}'\partial^{-1}u_{7}',\ l_{2,5}'=u_{1}'\partial^{-1}u_{5}'+u_{5}'\partial^{-1}u_{1}',\ l_{2,6}'=-u_{1}'\partial^{-1}u_{6}'+u_{10}'\partial^{-1}u_{8}'+u_{12}'\partial^{-1}u_{2}',\\& l_{2,7}'=\frac{1}{2}(u_{1}'\partial^{-1}u_{7}'-u_{5}'\partial^{-1}u_{10}'+u_{7}'\partial^{-1}u_{1}'-u_{10}'\partial^{-1}u_{5}'),\ l_{2,8}'=-\frac{1}{2}(u_{1}'\partial^{-1}u_{8}'+u_{10}'\partial^{-1}u_{3}'+u_{12}'\partial^{-1}u_{9}'),\\& l_{2,9}'=\frac{1}{2}(-u_{1}'\partial^{-1}u_{9}'+u_{5}'\partial^{-1}u_{8}'-u_{7}'\partial^{-1}u_{3}'),\  l_{2,10}'=\frac{1}{2}(u_{1}'\partial^{-1}u_{10}'+u_{7}'\partial^{-1}u_{12}'+u_{10}'\partial^{-1}u_{1}'+u_{12}'\partial^{-1}u_{7}'),\\& l_{2,11}'=-u_{1}'\partial^{-1}u_{11}'+u_{5}'\partial^{-1}u_{2}'-u_{7}'\partial^{-1}u_{9}',\ l_{2,12}'=u_{1}'\partial^{-1}u_{12}'+u_{12}'\partial^{-1}u_{1}',\
		l_{3,1}'=\frac{1}{2}(u_{7}'\partial^{-1}u_{10}'+u_{10}'\partial^{-1}u_{7}'),\\& l_{3,2}'=0,\  l_{3,3}'=\frac{1}{2}\partial-(u_{4}'\partial^{-1}u_{3}'+u_{10}'\partial^{-1}u_{9}'),\ l_{3,4}'=u_{4}'\partial^{-1}u_{4}',\ l_{3,5}'=-u_{7}'\partial^{-1}u_{7}',\ l_{3,6}'=0,\\& l_{3,7}'=\frac{1}{2}(u_{4}'\partial^{-1}u_{7}'+u_{7}'\partial^{-1}u_{4}'),\ l_{3,8}'=\frac{1}{2}(-u_{4}'\partial^{-1}u_{8}'+u_{7}'\partial^{-1}u_{6}'-u_{10}'\partial^{-1}u_{2}'),\\& l_{3,9}'=-\frac{1}{2}(u_{4}'\partial^{-1}u_{9}'+u_{7}'\partial^{-1}u_{2}'+u_{10}'\partial^{-1}u_{11}'),\  l_{3,10}'=\frac{1}{2}(u_{4}'\partial^{-1}u_{10}'+u_{10}'\partial^{-1}u_{4}'),\ l_{3,11}'=0,\\& l_{3,12}'=u_{10}'\partial^{-1}u_{10}',\
		l_{4,1}'=0,\ l_{4,2}'=-\frac{1}{2}(u_{8}'\partial^{-1}u_{9}'+u_{9}'\partial^{-1}u_{8}'),\ l_{4,3}'=-u_{3}'\partial^{-1}u_{3}',\\& l_{4,4}'=-\frac{1}{2}\partial+u_{3}'\partial^{-1}u_{4}'+u_{8}'\partial^{-1}u_{7}'+u_{9}'\partial^{-1}u_{10}',\ l_{4,5}'=0,\ l_{4,6}=u_{8}'\partial^{-1}u_{8}',\\& l_{4,7}'=\frac{1}{2}(u_{3}'\partial^{-1}u_{7}'-u_{8}'\partial^{-1}u_{5}'+u_{9}'\partial^{-1}u_{1}'),\ l_{4,8}'=-\frac{1}{2}(u_{3}'\partial^{-1}u_{8}'+u_{8}'\partial^{-1}u_{3}'),\ l_{4,9}'=-\frac{1}{2}(u_{3}'\partial^{-1}u_{9}'+u_{9}'\partial^{-1}u_{3}'),\\& l_{4,10}'=\frac{1}{2}(u_{3}'\partial^{-1}u_{10}'+u_{8}'\partial^{-1}u_{1}'+u_{9}'\partial^{-1}u_{12}'),\ l_{4,11}'=-u_{9}'\partial^{-1}u_{9}',\ l_{4,12}'=0,\\&
		l_{5,1}'=\frac{1}{2}(-u_{2}'\partial^{-1}u_{12}'+u_{6}'\partial^{-1}u_{1}'-u_{8}'\partial^{-1}u_{10}'),\ l_{5,2}'=-\frac{1}{2}(u_{2}'\partial^{-1}u_{6}'+u_{6}'\partial^{-1}u_{2}'),\ l_{5,3}=u_{8}'\partial^{-1}u_{8}',\ l_{5,4}'=0,\\& l_{5,5}'=-\frac{1}{2}\partial+u_{2}'\partial^{-1}u_{1}'+u_{6}'\partial^{-1}u_{5}'+u_{8}'\partial^{-1}u_{7}',\ l_{5,6}'=-u_{6}'\partial^{-1}u_{6}',\\& l_{5,7}'=\frac{1}{2}(-u_{2}'\partial^{-1}u_{10}'+u_{6}'\partial^{-1}u_{7}'-u_{8}'\partial^{-1}u_{4}'),\ l_{5,8}'=-\frac{1}{2}(u_{6}'\partial^{-1}u_{8}'+u_{8}'\partial^{-1}u_{6}'),\ l_{5,9}'=\frac{1}{2}(u_{2}'\partial^{-1}u_{8}'+u_{8}'\partial^{-1}u_{2}'),\\& l_{5,10}'=0,\ l_{5,11}'=u_{2}'\partial^{-1}u_{2}',\ l_{5,12}'=0,\ l_{6,1}'=\frac{1}{2}(u_{1}'\partial^{-1}u_{5}'+u_{5}'\partial^{-1}u_{1}'),\\&  l_{6,2}'=\frac{1}{2}(u_{1}'\partial^{-1}u_{11}'-u_{5}'\partial^{-1}u_{2}'+u_{7}'\partial^{-1}u_{9}'),\ l_{6,3}'=0,\ l_{6,4}'=-u_{7}'\partial^{-1}u_{7}',\ l_{6,5}'=u_{5}'\partial^{-1}u_{5}',\\& l_{6,6}'=\frac{1}{2}\partial-(u_{1}'\partial^{-1}u_{2}'+u_{5}'\partial^{-1}u_{6}'+u_{7}'\partial^{-1}u_{8}'),\ l_{6,7}'=\frac{1}{2}(u_{5}'\partial^{-1}u_{7}'+u_{7}'\partial^{-1}u_{5}'),\\& l_{6,8}'=\frac{1}{2}(u_{1}'\partial^{-1}u_{9}'-u_{5}'\partial^{-1}u_{8}'+u_{7}'\partial^{-1}u_{3}'),\ l_{6,9}'=0,\ l_{6,10}'=-\frac{1}{2}(u_{1}'\partial^{-1}u_{7}'+u_{7}'\partial^{-1}u_{1}'),\ l_{6,11}'=0,\\& l_{6,12}'=-u_{1}'\partial^{-1}u_{1}',\
		l_{7,1}'=\frac{1}{2}(u_{3}'\partial^{-1}u_{10}'+u_{8}'\partial^{-1}u_{1}'+u_{9}'\partial^{-1}u_{12}'),\\& l_{7,2}'=\frac{1}{2}(-u_{2}'\partial^{-1}u_{8}'+u_{6}'\partial^{-1}u_{9}'-u_{8}'\partial^{-1}u_{2}'+u_{9}'\partial^{-1}u_{6}'),\ l_{7,3}'=-(u_{3}'\partial^{-1}u_{8}'+u_{8}'\partial^{-1}u_{3}'),\\& l_{7,4}'=u_{2}'\partial^{-1}u_{10}'-u_{6}'\partial^{-1}u_{7}'+u_{8}'\partial^{-1}u_{4}',\ l_{7,5}'=-u_{3}'\partial^{-1}u_{7}'+u_{8}'\partial^{-1}u_{5}'-u_{9}'\partial^{-1}u_{1}',\\& l_{7,6}'=-(u_{6}'\partial^{-1}u_{8}'+u_{8}'\partial^{-1}u_{6}'),\ l_{7,7}'=\frac{1}{2}\partial+u_{8}'\partial^{-1}u_{7}'+\frac{1}{2}(u_{2}'\partial^{-1}u_{1}'+u_{3}'\partial^{-1}u_{4}'+u_{6}'\partial^{-1}u_{5}'+u_{9}'\partial^{-1}u_{10}'),\\& l_{7,8}'=-u_{8}'\partial^{-1}u_{8}'+\frac{1}{2}(u_{3}'\partial^{-1}u_{6}'+u_{6}'\partial^{-1}u_{3}'),\ l_{7,9}'=-\frac{1}{2}(u_{2}'\partial^{-1}u_{3}'+u_{3}'\partial^{-1}u_{2}'+u_{8}'\partial^{-1}u_{9}'+u_{9}'\partial^{-1}u_{8}'),\\& l_{7,10}'=\frac{1}{2}(u_{2}'\partial^{-1}u_{12}'-u_{6}'\partial^{-1}u_{1}'+u_{8}'\partial^{-1}u_{10}'),\ l_{7,11}'=-(u_{2}'\partial^{-1}u_{9}'+u_{9}'\partial^{-1}u_{2}'),\ l_{7,12}'=0,\\&
		l_{8,1}'=\frac{1}{2}(u_{1}'\partial^{-1}u_{7}'-u_{5}'\partial^{-1}u_{10}'+u_{7}'\partial^{-1}u_{1}'-u_{10}'\partial^{-1}u_{5}'),\ l_{8,2}'=-\frac{1}{2}(u_{4}'\partial^{-1}u_{9}'+u_{7}'\partial^{-1}u_{2}'+u_{10}'\partial^{-1}u_{11}'),\\& l_{8,3}'=-u_{1}'\partial^{-1}u_{9}'+u_{5}'\partial^{-1}u_{8}'-u_{7}'\partial^{-1}u_{3}',\ l_{8,4}'=u_{4}'\partial^{-1}u_{7}'+u_{7}'\partial^{-1}u_{4}',\ l_{8,5}'=u_{5}'\partial^{-1}u_{7}'+u_{7}'\partial^{-1}u_{5}',\\& l_{8,6}'=u_{4}'\partial^{-1}u_{8}'-u_{7}'\partial^{-1}u_{6}'+u_{10}'\partial^{-1}u_{2}',\  l_{8,7}'=u_{7}'\partial^{-1}u_{7}'-\frac{1}{2}(u_{4}'\partial^{-1}u_{5}'+u_{5}'\partial^{-1}u_{4}'),\\& l_{8,8}'=\frac{1}{2}\partial-u_{7}'\partial^{-1}u_{8}'-\frac{1}{2}(u_{1}'\partial^{-1}u_{2}'+u_{4}'\partial^{-1}u_{3}'+u_{5}'\partial^{-1}u_{6}'+u_{10}'\partial^{-1}u_{9}'),\\& l_{8,9}'=\frac{1}{2}(-u_{1}'\partial^{-1}u_{11}'+u_{5}'\partial^{-1}u_{2}'-u_{7}'\partial^{-1}u_{9}'),\  l_{8,10}'=\frac{1}{2}(u_{1}'\partial^{-1}u_{4}'+u_{4}'\partial^{-1}u_{1}'+u_{7}'\partial^{-1}u_{10}'+u_{10}'\partial^{-1}u_{7}'),\\& l_{8,11}'=0,\ l_{8,12}'=u_{1}'\partial^{-1}u_{10}'+u_{10}'\partial^{-1}u_{1}',\
		l_{9,1}'=\frac{1}{2}(u_{1}'\partial^{-1}u_{10}'+u_{7}'\partial^{-1}u_{12}'+u_{10}'\partial^{-1}u_{1}'+u_{12}'\partial^{-1}u_{7}'),\\& l_{9,2}'=\frac{1}{2}(-u_{4}'\partial^{-1}u_{8}'+u_{7}'\partial^{-1}u_{6}'-u_{10}'\partial^{-1}u_{2}'),\ l_{9,3}'=-(u_{1}'\partial^{-1}u_{8}'+u_{10}'\partial^{-1}u_{3}'+u_{12}'\partial^{-1}u_{9}'),\\& l_{9,4}'=u_{4}'\partial^{-1}u_{10}'+u_{10}'\partial^{-1}u_{4}',\  l_{9,5}'=-(u_{1}'\partial^{-1}u_{7}'+u_{7}'\partial^{-1}u_{1}'),\ l_{9,6}'=0,\\& l_{9,7}'=\frac{1}{2}(u_{1}'\partial^{-1}u_{4}'+u_{4}'\partial^{-1}u_{1}'+u_{7}'\partial^{-1}u_{10}'+u_{10}'\partial^{-1}u_{7}'),\ l_{9,8}'=\frac{1}{2}(u_{1}'\partial^{-1}u_{6}'-u_{10}'\partial^{-1}u_{8}'-u_{12}'\partial^{-1}u_{2}'),\\& l_{9,9}'=\frac{1}{2}\partial-u_{10}'\partial^{-1}u_{9}'-\frac{1}{2}(u_{1}'\partial^{-1}u_{2}'+u_{4}'\partial^{-1}u_{3}'+u_{7}'\partial^{-1}u_{8}'+u_{12}'\partial^{-1}u_{11}'),\\& l_{9,10}'=u_{10}'\partial^{-1}u_{10}'+\frac{1}{2}(u_{4}'\partial^{-1}u_{12}'+u_{12}'\partial^{-1}u_{4}'),\  l_{9,11}'=-(u_{4}'\partial^{-1}u_{9}'+u_{7}'\partial^{-1}u_{2}'+u_{10}'\partial^{-1}u_{11}'),\\& l_{9,12}'=u_{10}'\partial^{-1}u_{12}'+u_{12}'\partial^{-1}u_{10}',\
		l_{10,1}'=\frac{1}{2}(u_{3}'\partial^{-1}u_{7}'-u_{8}'\partial^{-1}u_{5}'+u_{9}'\partial^{-1}u_{1}'),\\& l_{10,2}'=-\frac{1}{2}(u_{2}'\partial^{-1}u_{9}'+u_{8}'\partial^{-1}u_{11}'+u_{9}'\partial^{-1}u_{2}'+u_{11}'\partial^{-1}u_{8}'),\ l_{10,3}'=-(u_{3}'\partial^{-1}u_{9}'+u_{9}'\partial^{-1}u_{3}'),\\& l_{10,4}'=u_{2}'\partial^{-1}u_{7}'+u_{9}'\partial^{-1}u_{4}'+u_{11}'\partial^{-1}u_{10}',\ l_{10,5}'=0,\ l_{10,6}'=u_{2}'\partial^{-1}u_{8}'+u_{8}'\partial^{-1}u_{2}',\\& l_{10,7}'=\frac{1}{2}(-u_{2}'\partial^{-1}u_{5}'+u_{9}'\partial^{-1}u_{7}'+u_{11}'\partial^{-1}u_{1}'),\ l_{10,8}'=-\frac{1}{2}(u_{2}'\partial^{-1}u_{3}'+u_{3}'\partial^{-1}u_{2}'+u_{8}'\partial^{-1}u_{9}'+u_{9}'\partial^{-1}u_{8}'),\\& l_{10,9}'=-u_{9}'\partial^{-1}u_{9}'-\frac{1}{2}(u_{3}'\partial^{-1}u_{11}'+u_{11}'\partial^{-1}u_{3}'),\\& l_{10,10}'=-\frac{1}{2}\partial+u_{9}'\partial^{-1}u_{10}'+\frac{1}{2}(u_{2}'\partial^{-1}u_{1}'+u_{3}'\partial^{-1}u_{4}'+u_{8}'\partial^{-1}u_{7}'+u_{11}'\partial^{-1}u_{12}'),\\& l_{10,11}'=-(u_{9}'\partial^{-1}u_{11}'+u_{11}'\partial^{-1}u_{9}'),\ l_{10,12}'=u_{3}'\partial^{-1}u_{10}'+u_{8}'\partial^{-1}u_{1}'+u_{9}'\partial^{-1}u_{12}',\\&
		l_{11,1}'=\frac{1}{2}(u_{1}'\partial^{-1}u_{12}'+u_{12}'\partial^{-1}u_{1}'),\ l_{11,2}'=\frac{1}{2}(u_{1}'\partial^{-1}u_{6}'-u_{10}'\partial^{-1}u_{8}'-u_{12}'\partial^{-1}u_{2}'),\ l_{11,3}'=0,\ l_{11,4}'=u_{10}'\partial^{-1}u_{10}',\\& l_{11,5}'=-u_{1}'\partial^{-1}u_{1}',\ l_{11,6}'=0,\ l_{11,7}'=\frac{1}{2}(u_{1}'\partial^{-1}u_{10}'+u_{10}'\partial^{-1}u_{1}'),\ l_{11,8}'=0,\\& l_{11,9}'=-\frac{1}{2}(u_{1}'\partial^{-1}u_{8}'+u_{10}'\partial^{-1}u_{3}'+u_{12}'\partial^{-1}u_{9}'),\ l_{11,10}'=\frac{1}{2}(u_{10}'\partial^{-1}u_{12}'+u_{12}'\partial^{-1}u_{10}'),\\& l_{11,11}'=\frac{1}{2}\partial-(u_{1}'\partial^{-1}u_{2}'+u_{10}'\partial^{-1}u_{9}'+u_{12}'\partial^{-1}u_{11}'),\ l_{11,12}'=u_{12}'\partial^{-1}u_{12}',\\&
		l_{12,1}'=\frac{1}{2}(-u_{2}'\partial^{-1}u_{5}'+u_{9}'\partial^{-1}u_{7}'+u_{11}'\partial^{-1}u_{1}'),\  l_{12,2}'=-\frac{1}{2}(u_{2}'\partial^{-1}u_{11}'+u_{11}'\partial^{-1}u_{2}'),\ l_{12,3}'=-u_{9}'\partial^{-1}u_{9}',\\& l_{12,4}'=0,\ l_{12,5}'=0,\ l_{12,6}'=u_{2}'\partial^{-1}u_{2}',\ l_{12,7}=0,\ l_{12,8}'=-\frac{1}{2}(u_{2}'\partial^{-1}u_{9}'+u_{9}'\partial^{-1}u_{2}'),\\& l_{12,9}'=-\frac{1}{2}(u_{9}'\partial^{-1}u_{11}'+u_{11}'\partial^{-1}u_{9}'),\ l_{12,10}'=\frac{1}{2}(u_{2}'\partial^{-1}u_{7}'+u_{9}'\partial^{-1}u_{4}'+u_{11}'\partial^{-1}u_{10}'),\ l_{12,11}'=-u_{11}'\partial^{-1}u_{11}',\\& l_{12,12}'=-\frac{1}{2}\partial+u_{2}'\partial^{-1}u_{1}'+u_{9}'\partial^{-1}u_{10}'+u_{11}'\partial^{-1}u_{12}'
	\end{align*}
	To construct the Hamiltonian structure, we employ trace identities \cite{ref13}, and have
	\begin{align*}
		&\left\langle V_{0},\frac{\partial U_{1}}{\partial\lambda}\right\rangle = 2a-2b-2c ,\
		\left\langle V_{0},\frac{\partial U_{1}}{\partial u_{1}'}\right\rangle = 2g ,\
		\left\langle V_{0},\frac{\partial U_{1}}{\partial u_{2}'}\right\rangle = 2f ,\
		\left\langle V_{0},\frac{\partial U_{1}}{\partial u_{3}'}\right\rangle = o ,\\
		&\left\langle V_{0},\frac{\partial U_{1}}{\partial u_{4}'}\right\rangle = m ,\
		\left\langle V_{0},\frac{\partial U_{1}}{\partial u_{5}'}\right\rangle = q ,\
		\left\langle V_{0},\frac{\partial U_{1}}{\partial u_{6}'}\right\rangle = p ,\
		\left\langle V_{0},\frac{\partial U_{1}}{\partial u_{7}'}\right\rangle = 2e ,\
		\left\langle V_{0},\frac{\partial U_{1}}{\partial u_{8}'}\right\rangle = 2d ,\\
		&\left\langle V_{0},\frac{\partial U_{1}}{\partial u_{9}'}\right\rangle = 2u ,\
		\left\langle V_{0},\frac{\partial U_{1}}{\partial u_{10}'}\right\rangle = 2t ,\
		\left\langle V_{0},\frac{\partial U_{1}}{\partial u_{11}'}\right\rangle = w ,\
		\left\langle V_{0},\frac{\partial U_{1}}{\partial u_{12}'}\right\rangle = v.
	\end{align*}
	Substituting the above formulate into the trace identity yields
	\begin{align*}
		\frac{\delta}{\delta u'}(2a-2b-2c)=\lambda^{-\tau}\frac{\partial}{\partial\lambda}\lambda^{\tau}
		\left(\begin{array}{c}
			2g\\
			2f\\
			o\\
			m\\
			q\\
			p\\
			2e\\
			2d\\
			2u\\
			2t\\
			w\\
			v
		\end{array}\right),
	\end{align*}
	where $\tau=\frac{\lambda}{2}\frac{d}{dx}ln|tr(V_{0}^{2})|$. Balancing coefficients of each power of  in the above equality gives rise to
	\begin{align*}
		\frac{\delta}{\delta u'}(2a_{n+1}-2b_{n+1}-2c_{n+1})=(\tau-n)
		\left(\begin{array}{c}
			2g_{n}\\
			2f_{n}\\
			o_{n}\\
			m_{n}\\
			q_{n}\\
			p_{n}\\
			2e_{n}\\
			2d_{n}\\
			2u_{n}\\
			2t_{n}\\
			w_{n}\\
			v_{n}
		\end{array}\right),
	\end{align*}
	Taking $n=1$, gives $\tau=-1$. Therefore we establish the following equation:
	\begin{align*}
		P_{2,n+1}=
		\left(\begin{array}{c}
			2g_{n+1}\\
			2f_{n+1}\\
			o_{n+1}\\
			m_{n+1}\\
			q_{n+1}\\
			p_{n+1}\\
			2e_{n+1}\\
			2f_{n+1}\\
			2u_{n+1}\\
			2t_{n+1}\\
			w_{n+1}\\
			v_{n+1}
		\end{array}\right)=\frac{\delta}{\delta u'}((\frac{-2}{n+2})(a_{n+2}-b_{n+2}-c_{n+2})).
	\end{align*}
	Thus, we see:
	\begin{align*}
		u_{t}'=J_{2}P_{2,n+1}=J_{2}\frac{\delta H_{n+1}^{2}}{\delta u},\  H_{n+1}^{2}=(\frac{-2}{n+2})(a_{n+2}-b_{n+2}-c_{n+2}),\ n\geq0.
	\end{align*}
	It is said that the hierarchy (\ref{eq:2.8}) have the Hamiltonian structure, and it is easy to verity that $J_{2}L_{2}=L_{2}^{*}J_{2}$. Therefore, the hierarchy (\ref{eq:2.8}) is Liouville integrable.
	
	When $n=1$, the hierarchy (\ref{eq:2.8}) reduces to the first integrable system
	\begin{align*}
		u_{1t}'=&\frac{1}{2}u_{1x}'(\alpha-\gamma)-\frac{1}{2}u_{7}'\partial^{-1}(u_{1}'u_{8}'+u_{3}'u_{10}'+u_{9}'u_{12}')(\beta-\gamma)+\frac{1}{2}u_{12}'\partial^{-1}(-u_{1}'u_{11}'+u_{2}'u_{5}'-u_{7}'u_{9}')(\alpha+\gamma)\\
		&-\frac{1}{2}u_{5}'\partial^{-1}(-u_{1}'u_{6}'+u_{2}'u_{12}'+u_{8}'u_{10}')(\alpha+\gamma)+\frac{1}{2}u_{10}'\partial^{-1}(-u_{1}'u_{9}'-u_{3}'u_{7}'+u_{5}'u_{8}')(\alpha+\beta),\\
		u_{2t}'=&\frac{1}{2}u_{2x}'(\alpha-\gamma)-\frac{1}{2}u_{8}'\partial^{-1}(u_{2}'u_{7}'+u_{4}'u_{9}'+u_{10}'u_{11}')(\beta-\gamma)+\frac{1}{2}u_{6}'\partial^{-1}(-u_{1}'u_{11}'+u_{2}'u_{5}'-u_{7}'u_{9}')(\alpha+\gamma)\\
		&-\frac{1}{2}u_{11}'\partial^{-1}(-u_{1}'u_{6}'+u_{2}'u_{12}'+u_{8}'u_{10}')(\alpha+\gamma)-\frac{1}{2}u_{9}'\partial^{-1}(u_{2}'u_{10}'+u_{4}'u_{8}'-u_{6}'u_{7}')(\alpha+\beta),\\
		u_{3t}'=&-u_{3x}'\beta+\frac{1}{2}u_{8}'\partial^{-1}(-u_{1}'u_{9}'-u_{3}'u_{7}'+u_{5}'u_{8}')(\alpha+\beta)+u_{9}'\partial^{-1}(u_{1}'u_{8}'+u_{3}'u_{10}'+u_{9}'u_{12}')(\beta-\gamma),\\
		u_{4t}'=&-u_{4x}'\beta+\frac{1}{2}u_{10}'\partial^{-1}(u_{2}'u_{7}'+u_{4}'u_{9}'+u_{10}'u_{11}')(\beta-\gamma)+u_{7}'\partial^{-1}(u_{2}'u_{10}'+u_{4}'u_{8}'-u_{6}'u_{7}')(\alpha+\beta),\\
		u_{5t}'=&u_{5x}'\alpha-u_{1}'\partial^{-1}(-u_{1}'u_{11}'+u_{2}'u_{5}'-u_{7}'u_{9}')(\alpha+\gamma)-u_{7}'\partial^{-1}(-u_{1}'u_{9}'-u_{3}'u_{7}'+u_{5}'u_{8}')(\alpha+\beta),\\
		u_{6t}'=&u_{6x}'\alpha-\frac{1}{2}u_{2}'\partial^{-1}(-u_{1}'u_{6}'+u_{2}'u_{12}'+u_{8}'u_{10}')(\alpha+\gamma)+u_{8}'\partial^{-1}(u_{2}'u_{10}'+u_{4}'u_{8}'-u_{6}'u_{7}')(\alpha+\beta),\\
		u_{7t}'=&\frac{1}{2}u_{7x}'(\alpha-\beta)+\frac{1}{2}u_{1}'\partial^{-1}(u_{2}'u_{7}'+u_{4}'u_{9}'+u_{10}'u_{11}')(\beta-\gamma)+\frac{1}{2}u_{10}'\partial^{-1}(-u_{1}'u_{11}'+u_{2}'u_{5}'-u_{7}'u_{9}')(\alpha+\gamma)\\
		&+\frac{1}{2}u_{4}'\partial^{-1}(-u_{1}'u_{9}'-u_{3}'u_{7}'+u_{5}'u_{8}')(\alpha+\beta)-\frac{1}{2}u_{5}'\partial^{-1}(u_{2}'u_{10}'+u_{4}'u_{8}'-u_{6}'u_{7}')(\alpha+\beta),\\
		u_{8t}'=&\frac{1}{2}u_{8x}'(\alpha-\beta)+\frac{1}{2}u_{2}'\partial^{-1}(u_{1}'u_{8}'+u_{3}'u_{10}'+u_{9}'u_{12}')(\beta-\gamma)-\frac{1}{2}u_{9}'\partial^{-1}(-u_{1}'u_{6}'+u_{2}'u_{12}'+u_{8}'u_{10}')(\alpha+\gamma)\\
		&+\frac{1}{2}u_{4}'\partial^{-1}(-u_{1}'u_{9}'-u_{3}'u_{7}'+u_{5}'u_{8}')(\alpha+\beta)-\frac{1}{2}u_{3}'\partial^{-1}(u_{2}'u_{10}'+u_{4}'u_{8}'-u_{6}'u_{7}')(\alpha+\beta),\\
		u_{9t}'=&-\frac{1}{2}u_{9x}'(\beta+\gamma)-\frac{1}{2}u_{8}'\partial^{-1}(-u_{1}'u_{11}'+u_{2}'u_{5}'-u_{7}'u_{9}')(\alpha+\gamma)+\frac{1}{2}u_{11}'\partial^{-1}(u_{1}'u_{8}'+u_{3}'u_{10}'+u_{9}'u_{12}')(\beta-\gamma),\\
		&-\frac{1}{2}u_{3}'\partial^{-1}(u_{2}'u_{7}'+u_{4}'u_{9}'+u_{10}'u_{11}')(\beta-\gamma)-\frac{1}{2}u_{2}'\partial^{-1}(-u_{1}'u_{9}'-u_{3}'u_{7}'+u_{5}'u_{8}')(\alpha+\beta),\\
		u_{10t}'=&-\frac{1}{2}u_{10x}'(\beta+\gamma)-\frac{1}{2}u_{4}'\partial^{-1}(u_{1}'u_{8}'+u_{3}'u_{10}'+u_{9}'u_{12}')(\beta-\gamma)+\frac{1}{2}u_{12}'\partial^{-1}(u_{2}'u_{7}'+u_{4}'u_{9}'+u_{10}'u_{11}')(\beta-\gamma),\\
		&+\frac{1}{2}u_{7}'\partial^{-1}(-u_{1}'u_{6}'+u_{2}'u_{12}'+u_{8}'u_{10}')(\alpha+\gamma)+\frac{1}{2}u_{1}'\partial^{-1}(u_{2}'u_{10}'+u_{4}'u_{8}'-u_{6}'u_{7}')(\alpha+\beta),\\
		u_{11t}'=&-u_{11x}'\gamma-u_{2}'\partial^{-1}(-u_{1}'u_{11}'+u_{2}'u_{5}'-u_{7}'u_{9}')(\alpha+\gamma)-u_{9}'\partial^{-1}(u_{2}'u_{7}'+u_{4}'u_{9}'+u_{10}'u_{11}')(\beta-\gamma),\\
		u_{12t}'=&-u_{12x}'\gamma-u_{10}'\partial^{-1}(u_{1}'u_{8}'+u_{3}'u_{10}'+u_{9}'u_{12}')(\beta-\gamma)+u_{1}'\partial^{-1}(-u_{1}'u_{6}'+u_{2}'u_{12}'+u_{8}'u_{10}')(\alpha+\gamma).
	\end{align*}
	Following the approach in Section 2.1, we construct a second integrable soliton hierarchy by selecting different spectral matrices, and similarly present its corresponding integrable system for the case $n=1$ as a concrete example.
	
	\section{ Integrable Coupling Systems for the Hierarchy of $\mathfrak{sp}(6)$ }\label{Sec:3}
	In this section, we will use Kronecker product to construct integrable coupling systems of Lie algebra $\mathfrak{sp}(6)$.
	
	Let $U_{3}$ and $V_{3}$ have the forms
	\begin{align*}
		&U_{3}=\left(
		\begin{array}{cc}
			1&0\\
			0&1
		\end{array}\right)\otimes U_0+
		\left(
		\begin{array}{cc}
			0&1\\
			0&0
		\end{array}\right)\otimes U_2,\\
		&V_{3}=\left(
		\begin{array}{cc}
			1&0\\
			0&1
		\end{array}\right)\otimes V_0+
		\left(
		\begin{array}{cc}
			0&1\\
			0&0
		\end{array}\right)\otimes V_2,
	\end{align*}
	where $U_{0}$, $V_{0}$ be defined as (\ref{eq:2.2}), (\ref{eq:2.3}), and $U_{2}$, $V_{2}\in\widetilde{\mathfrak{sp}(6)}$ be defined as follow:
	\begin{align*}
		U_{2}=\left(
		\begin{array}{cccccc}
			0&0&0&u_{5}^{*}&u_{7}^{*}&u_{1}^{*}\\
			0&0&0&u_{7}^{*}&u_{3}^{*}&u_{9}^{*}\\
			0&0&0&u_{1}^{*}&u_{9}^{*}&u_{11}^{*}\\
			u_{6}^{*}&u_{8}^{*}&u_{2}^{*}&0&0&0\\
			u_{8}^{*}&u_{4}^{*}&u_{10}^{*}&0&0&0\\
			u_{2}^{*}&u_{10}^{*}&u_{12}^{*}&0&0&0
		\end{array}\right),
	\end{align*}
	\begin{align*}
		V_{2}=\left(
		\begin{array}{cccccc}
			a^{*}&d^{*}&f^{*}&p^{*}&r^{*}&k^{*}\\
			e^{*}&b^{*}&h^{*}&r^{*}&m^{*}&t^{*}\\
			g^{*}&j^{*}&c^{*}&k^{*}&t^{*}&v^{*}\\
			q^{*}&s^{*}&l^{*}&-a^{*}&-e^{*}&-g^{*}\\
			s^{*}&o^{*}&u^{*}&-d^{*}&-b^{*}&-j^{*}\\
			l^{*}&u^{*}&w^{*}&-f^{*}&-h^{*}&-c^{*}\\
		\end{array}\right)=\sum_{i\geq0}\left(
		\begin{array}{cccccc}
			a^{*}_{i}&d^{*}_{i}&f^{*}_{i}&p^{*}_{i}&r^{*}_{i}&k^{*}_{i}\\
			e^{*}_{i}&b^{*}_{i}&h^{*}_{i}&r^{*}_{i}&m^{*}_{i}&t^{*}_{i}\\
			g^{*}_{i}&j^{*}_{i}&c^{*}_{i}&k^{*}_{i}&t^{*}_{i}&v^{*}_{i}\\
			q^{*}_{i}&s^{*}_{i}&l^{*}_{i}&-a^{*}_{i}&-e^{*}_{i}&-g^{*}_{i}\\
			s^{*}_{i}&o^{*}_{i}&u^{*}_{i}&-d^{*}_{i}&-b^{*}_{i}&-j^{*}_{i}\\
			l^{*}_{i}&u^{*}_{i}&w^{*}_{i}&-f^{*}_{i}&-h^{*}_{i}&-c^{*}_{i}.
		\end{array}\right)
	\end{align*}
	Then we have a new pair of $U_{3}$ and $V_{3}$ as follow:
	\begin{align*}
		U_{3}=\left(\begin{array}{cc}
			U_{0}&U_{2}\\
			0&U_{0}
		\end{array}\right),\quad
		V_{3}=\left(\begin{array}{cc}
			V_{0}&V_{2}\\
			0&V_{0}
		\end{array}\right).
	\end{align*}
	
	Therefore, the stationary zero curvature representation $V_{3,x}=[U_{3},V_{3}]$ is equivalent to
	\begin{align*}
		\begin{cases}
			V_{0,x}=[U_{0},V_{0}],\\
			V_{3,x}=[U_{0},V_{2}]+[U_{2},V_{0}],
		\end{cases}
	\end{align*}
	and the corresponding enlarged zero curvature equation $U_{3,t}-V_{3,x}+[U_{3},V_{3}]=0$ is equivalent to
	\begin{align*}
		\begin{cases}
			U_{0,t}-V_{0,x}+[U_{0},V_{0}]=0,\\
			U_{2,t}-V_{2,x}+[U_{0},V_{2}]+[U_{2},V_{0}]=0.
		\end{cases}
	\end{align*}

	From the stationary zero curvature equation $V_{3,x}=[U_{3},V_{3}]$, we obtain (\ref{eq:2.4}) and
	\begin{align}
		\begin{cases}
			a_{x}^{*}=u_{1}l^{*}-u_{2}k^{*}+u_{5}q^{*}-u_{6}p^{*}+u_{7}s^{*}-u_{8}r^{*}+u_{1}^{*}l-u_{2}^{*}k+u_{5}^{*}q-u_{6}^{*}p+u_{7}^{*}s-u_{8}^{*}r,\\
			b_{x}^{*}=u_{3}o^{*}-u_{4}m^{*}+u_{7}s^{*}-u_{8}r^{*}+u_{9}u^{*}-u_{10}t^{*}+u_{3}^{*}o-u_{4}^{*}m+u_{7}^{*}s-u_{8}^{*}r+u_{9}^{*}u-u_{10}^{*}t,\\
			c_{x}^{*}=u_{1}l^{*}-u_{2}k^{*}+u_{9}u^{*}-u_{10}t^{*}+u_{11}w^{*}-u_{12}v^{*}+u_{1}^{*}l-u_{2}^{*}k+u_{9}^{*}u-u_{10}^{*}t+u_{11}^{*}w-u_{12}^{*}v,\\
			d_{x}^{*}=u_{1}u^{*}-u_{4}r^{*}+u_{5}s^{*}+u_{7}o^{*}-u_{8}p^{*}-u_{10}k^{*}+u_{1}^{*}u-u_{4}^{*}r+u_{5}^{*}s+u_{7}^{*}o-u_{8}^{*}p-u_{10}^{*}k,\\
			e_{x}^{*}=-u_{2}t^{*}+u_{3}s^{*}-u_{6}r^{*}+u_{7}q^{*}-u_{8}m^{*}+u_{9}l^{*}-u_{2}^{*}t+u_{3}^{*}s-u_{6}^{*}r+u_{7}^{*}q-u_{8}^{*}m+u_{9}^{*}l,\\
			f_{x}^{*}=u_{1}w^{*}-u_{2}p^{*}+u_{5}l^{*}+u_{7}u^{*}-u_{10}r^{*}-u_{12}k^{*}+u_{1}^{*}w-u_{2}^{*}p+u_{5}^{*}l+u_{7}^{*}u-u_{10}^{*}r-u_{12}^{*}k,\\
			g_{x}^{*}=u_{1}q^{*}-u_{2}v^{*}-u_{6}k^{*}-u_{8}t^{*}+u_{9}s^{*}+u_{11}l^{*}+u_{1}^{*}q-u_{2}^{*}v-u_{6}^{*}k-u_{8}^{*}t+u_{9}^{*}s+u_{11}^{*}l,\\
			h_{x}^{*}=-u_{2}r^{*}+u_{3}u^{*}+u_{7}l^{*}+u_{9}w^{*}-u_{10}m^{*}-u_{12}t^{*}-u_{2}^{*}r+u_{3}^{*}u+u_{7}^{*}l+u_{9}^{*}w-u_{10}^{*}m-u_{12}^{*}t,\\
			j_{x}^{*}=u_{1}s^{*}-u_{4}t^{*}-u_{8}k^{*}+u_{9}o^{*}-u_{10}v^{*}+u_{11}u^{*}+u_{1}^{*}s-u_{4}^{*}t-u_{8}^{*}k+u_{9}^{*}o-u_{10}^{*}v+u_{11}^{*}u,\\
			k_{x}^{*}=2\lambda k^{*}-u_{1}a^{*}-u_{1}c^{*}-u_{5}g^{*}-u_{7}j^{*}-u_{9}d^{*}-u_{11}f^{*}-u_{1}^{*}a-u_{1}^{*}c-u_{5}^{*}g-u_{7}^{*}j-u_{9}^{*}d-u_{11}^{*}f,\\
			l_{x}^{*}=-2\lambda l^{*}+u_{2}a^{*}+u_{2}c^{*}+u_{6}f^{*}+u_{8}h^{*}+u_{10}e^{*}+u_{12}g^{*}+u_{2}^{*}a+u_{2}^{*}c+u_{6}^{*}f+u_{8}^{*}h+u_{10}^{*}e+u_{12}^{*}g,\\
			m_{x}^{*}=2\lambda m^{*}-2u_{3}b^{*}-2u_{7}e^{*}-2u_{9}h^{*}-2u_{3}^{*}b-2u_{7}^{*}e-2u_{9}^{*}h,\\
			o_{x}^{*}=-2\lambda o^{*}+2u_{4}b^{*}+2u_{8}d^{*}+2u_{10}j^{*}+2u_{4}^{*}b+2u_{8}^{*}d+2u_{10}^{*}j,\\
			p_{x}^{*}=2\lambda p^{*}-2u_{1}f^{*}-2u_{5}a^{*}-2u_{7}d^{*}-2u_{1}^{*}f-2u_{5}^{*}a-2u_{7}^{*}d,\\
			q_{x}^{*}=-2\lambda q^{*}+2u_{2}g^{*}+2u_{6}a^{*}+2u_{8}e^{*}+2u_{2}^{*}g+2u_{6}^{*}a+2u_{8}^{*}e,\\
			r_{x}^{*}=2\lambda r^{*}-u_{1}h^{*}-u_{3}d^{*}-u_{5}e^{*}-u_{7}a^{*}-u_{7}b^{*}-u_{9}f^{*}-u_{1}^{*}h-u_{3}^{*}d-u_{5}^{*}e-u_{7}^{*}a-u_{7}^{*}b-u_{9}^{*}f,\\
			s_{x}^{*}=-2\lambda s^{*}+u_{2}j^{*}+u_{4}e^{*}+u_{6}d^{*}+u_{8}a^{*}+u_{8}b^{*}+u_{10}g^{*}+u_{2}^{*}j+u_{4}^{*}e+u_{6}^{*}d+u_{8}^{*}a+u_{8}^{*}b+u_{10}^{*}g,\\
			t_{x}^{*}=2\lambda t^{*}-u_{1}e^{*}-u_{3}j^{*}-u_{7}g^{*}-u_{9}b^{*}-u_{9}c^{*}-u_{11}h^{*}-u_{1}^{*}e-u_{3}^{*}j-u_{7}^{*}g-u_{9}^{*}b-u_{9}^{*}c-u_{11}^{*}h,\\
			u_{x}^{*}=-2\lambda u^{*}+u_{2}d^{*}+u_{4}h^{*}+u_{8}f^{*}+u_{10}b^{*}+u_{10}c^{*}+u_{12}j^{*}+u_{2}^{*}d+u_{4}^{*}h+u_{8}^{*}f+u_{10}^{*}b+u_{10}^{*}c+u_{12}^{*}j,\\
			v_{x}^{*}=2\lambda v^{*}-2u_{1}g^{*}-2u_{9}j^{*}-2u_{11}c^{*}-2u_{1}^{*}g-2u_{9}^{*}j-2u_{11}^{*}c,\\
			w_{x}^{*}=-2\lambda w^{*}+2u_{2}f^{*}+2u_{10}h^{*}+2u_{12}c^{*}+2u_{2}^{*}f+2u_{10}^{*}h+2u_{12}^{*}c.
		\end{cases}\label{eq:3.1}
	\end{align}
	Similarly, take the initial values
	\begin{align*}
		a_{0}^{*}=\alpha^{*}, b_{0}^{*}=\beta^{*}, c_{0}^{*}=\gamma^{*}, d_{0}^{*}=e_{0}^{*}=\dots=v_{0}^{*}=w_{0}^{*}=0.
	\end{align*}
	From (\ref{eq:3.1}), we have
	\begin{align*}
		a_{1}^{*}=&b_{1}^{*}=c_{1}^{*}=0,\\ d_{1}^{*}=&\frac{1}{2}\partial^{-1}(u_{1}u_{10}+u_{4}u_{7}+u_{5}u_{8})(\beta^{*}-\alpha^{*})+\frac{1}{2}\partial^{-1}(u_{1}u_{10}^{*}+u_{4}u_{7}^{*}+u_{5}u_{8}^{*})(\beta-\alpha)\\&+\frac{1}{2}\partial^{-1}(u_{1}^{*}u_{10}+u_{4}^{*}u_{7}+u_{5}^{*}u_{8})(\beta-\alpha),\\
		e_{1}^{*}=&\frac{1}{2}\partial^{-1}(u_{2}u_{9}+u_{3}u_{8}+u_{6}u_{7})(\alpha^{*}-\beta^{*})+\frac{1}{2}\partial^{-1}(u_{2}u_{9}^{*}+u_{3}u_{8}^{*}+u_{6}u_{7}^{*})(\alpha-\beta)\\&+\frac{1}{2}\partial^{-1}(u_{2}^{*}u_{9}+u_{3}^{*}u_{8}+u_{6}^{*}u_{7})(\alpha-\beta),\\
		f_{1}^{*}=&\frac{1}{2}\partial^{-1}(u_{1}u_{12}+u_{2}u_{5}+u_{7}u_{10})(\gamma^{*}-\alpha^{*})+\frac{1}{2}\partial^{-1}(u_{1}u_{12}^{*}+u_{2}u_{5}^{*}+u_{7}u_{10}^{*})(\gamma-\alpha)\\&+\frac{1}{2}\partial^{-1}(u_{1}^{*}u_{12}+u_{2}^{*}u_{5}+u_{7}^{*}u_{10})(\gamma-\alpha),\\
		g_{1}^{*}=&\frac{1}{2}\partial^{-1}(u_{1}u_{6}+u_{2}u_{11}+u_{8}u_{9})(\alpha^{*}-\gamma^{*})+\frac{1}{2}\partial^{-1}(u_{1}u_{6}^{*}+u_{2}u_{11}^{*}+u_{8}u_{9}^{*})(\alpha-\gamma)\\&+\frac{1}{2}\partial^{-1}(u_{1}^{*}u_{6}+u_{2}^{*}u_{11}+u_{8}^{*}u_{9})(\alpha-\gamma),\\
		h_{1}^{*}=&\frac{1}{2}\partial^{-1}(u_{2}u_{7}+u_{3}u_{10}+u_{9}u_{12})(\gamma^{*}-\beta^{*})+\frac{1}{2}\partial^{-1}(u_{2}u_{7}^{*}+u_{3}u_{10}^{*}+u_{9}u_{12}^{*})(\gamma-\beta)\\&+\frac{1}{2}\partial^{-1}(u_{2}^{*}u_{7}+u_{3}^{*}u_{10}+u_{9}^{*}u_{12})(\gamma-\beta),\\
		j_{1}^{*}=&\frac{1}{2}\partial^{-1}(u_{1}u_{8}+u_{4}u_{9}+u_{10}u_{11})(\beta^{*}-\gamma^{*})+\frac{1}{2}\partial^{-1}(u_{1}u_{8}^{*}+u_{4}u_{9}^{*}+u_{10}u_{11}^{*})(\beta-\gamma)\\&+\frac{1}{2}\partial^{-1}(u_{1}^{*}u_{8}+u_{4}^{*}u_{9}+u_{10}^{*}u_{11})(\beta-\gamma),\\
		k_{1}^{*}=&\frac{1}{2}u_{1}(\alpha^{*}+\gamma^{*})+\frac{1}{2}u_{1}^{*}(\alpha+\gamma),\
		l_{1}^{*}=\frac{1}{2}u_{2}(\alpha^{*}+\gamma^{*})+\frac{1}{2}u_{2}^{*}(\alpha+\gamma),\ m_{1}^{*}=u_{3}\beta^{*}+u_{3}^{*}\beta,\
		o_{1}^{*}=u_{4}\beta^{*}+u_{4}^{*}\beta,\\
		p_{1}^{*}=&u_{5}\alpha^{*}+u_{5}^{*}\alpha,\
		q_{1}^{*}=u_{6}\alpha^{*}+u_{6}^{*}\alpha,\ r_{1}^{*}=\frac{1}{2}u_{7}(\alpha^{*}+\beta^{*})+\frac{1}{2}u_{7}^{*}(\alpha+\beta),\
		s_{1}^{*}=\frac{1}{2}u_{8}(\alpha^{*}+\beta^{*})+\frac{1}{2}u_{8}^{*}(\alpha+\beta),\\t_{1}^{*}=&\frac{1}{2}u_{9}(\beta^{*}+\gamma^{*})+\frac{1}{2}u_{9}^{*}(\beta+\gamma),\ u_{1}^{*}=\frac{1}{2}u_{10}(\beta^{*}+\gamma^{*})+\frac{1}{2}u_{10}^{*}(\beta+\gamma),\ v_{1}^{*}=u_{11}\gamma^{*}+u_{11}^{*}\gamma,\
		w_{1}^{*}=u_{12}\gamma^{*}+u_{12}^{*}\gamma,\\
		k_{2}^{*}=&\frac{1}{4}u_{1x}(\alpha^{*}+\gamma^{*})+\frac{1}{4}u_{1x}^{*}(\alpha+\gamma)+\frac{1}{4}u_{5}\partial^{-1}(u_{1}u_{6}+u_{2}u_{11}+u_{8}u_{9})(\alpha^{*}-\gamma^{*})\\&+\frac{1}{4}u_{5}\partial^{-1}(u_{1}u_{6}^{*}+u_{2}u_{11}^{*}+u_{8}u_{9}^{*})(\alpha-\gamma)+\frac{1}{4}u_{5}\partial^{-1}(u_{1}^{*}u_{6}+u_{2}^{*}u_{11}+u_{8}^{*}u_{9})(\alpha-\gamma)\\&+\frac{1}{4}u_{5}^{*}\partial^{-1}(u_{1}u_{6}+u_{2}u_{11}+u_{8}u_{9})(\alpha-\gamma)+\frac{1}{4}u_{7}\partial^{-1}(u_{1}u_{8}+u_{4}u_{9}+u_{10}u_{11})(\beta^{*}-\gamma^{*})\\&+\frac{1}{4}u_{7}\partial^{-1}(u_{1}u_{8}^{*}+u_{4}u_{9}^{*}+u_{10}u_{11}^{*})(\beta-\gamma)+\frac{1}{4}u_{7}\partial^{-1}(u_{1}^{*}u_{8}+u_{4}^{*}u_{9}+u_{10}^{*}u_{11})(\beta-\gamma)\\&+\frac{1}{4}u_{7}^{*}\partial^{-1}(u_{1}u_{8}+u_{4}u_{9}+u_{10}u_{11})(\beta-\gamma)+\frac{1}{4}u_{9}\partial^{-1}(u_{1}u_{10}+u_{4}u_{7}+u_{5}u_{8})(\beta^{*}-\alpha^{*})\\&+\frac{1}{4}u_{9}\partial^{-1}(u_{1}u_{10}^{*}+u_{4}u_{7}^{*}+u_{5}u_{8}^{*})(\beta-\alpha)+\frac{1}{4}u_{9}\partial^{-1}(u_{1}^{*}u_{10}+u_{4}^{*}u_{7}+u_{5}^{*}u_{8})(\beta-\alpha)\\&+\frac{1}{4}u_{9}^{*}\partial^{-1}(u_{1}u_{10}+u_{4}u_{7}+u_{5}u_{8})(\beta-\alpha)+\frac{1}{4}u_{11}\partial^{-1}(u_{1}u_{12}+u_{2}u_{5}+u_{7}u_{10})(\gamma^{*}-\alpha^{*})\\&+\frac{1}{4}u_{11}\partial^{-1}(u_{1}u_{12}^{*}+u_{2}u_{5}^{*}+u_{7}u_{10}^{*})(\gamma-\alpha)+\frac{1}{4}u_{11}\partial^{-1}(u_{1}^{*}u_{12}+u_{2}^{*}u_{5}+u_{7}^{*}u_{10})(\gamma-\alpha)\\&+\frac{1}{4}u_{11}^{*}\partial^{-1}(u_{1}u_{12}+u_{2}u_{5}+u_{7}u_{10})(\gamma-\alpha),\\
		l_{2}^{*}=&-\frac{1}{4}u_{2x}(\alpha^{*}+\gamma^{*})-\frac{1}{4}u_{2x}^{*}(\alpha+\gamma)+\frac{1}{4}u_{6}\partial^{-1}(u_{1}u_{12}+u_{2}u_{5}+u_{7}u_{10})(\gamma^{*}-\alpha^{*})\\&+\frac{1}{4}u_{6}\partial^{-1}(u_{1}u_{12}^{*}+u_{2}u_{5}^{*}+u_{7}u_{10}^{*})(\gamma-\alpha)+\frac{1}{4}u_{6}\partial^{-1}(u_{1}^{*}u_{12}+u_{2}^{*}u_{5}+u_{7}^{*}u_{10})(\gamma-\alpha)\\&+\frac{1}{4}u_{6}^{*}\partial^{-1}(u_{1}u_{12}+u_{2}u_{5}+u_{7}u_{10})(\gamma-\alpha)+\frac{1}{4}u_{8}\partial^{-1}(u_{2}u_{7}+u_{3}u_{10}+u_{9}u_{12})(\gamma^{*}-\beta^{*})\\&+\frac{1}{4}u_{8}\partial^{-1}(u_{2}u_{7}^{*}+u_{3}u_{10}^{*}+u_{9}u_{12}^{*})(\gamma-\beta)+\frac{1}{4}u_{8}\partial^{-1}(u_{2}^{*}u_{7}+u_{3}^{*}u_{10}+u_{9}^{*}u_{12})(\gamma-\beta)\\&+\frac{1}{4}u_{8}^{*}\partial^{-1}(u_{2}u_{7}+u_{3}u_{10}+u_{9}u_{12})(\gamma-\beta)+\frac{1}{4}u_{10}\partial^{-1}(u_{2}u_{9}+u_{3}u_{8}+u_{6}u_{7})(\alpha^{*}-\beta^{*})\\&+\frac{1}{4}u_{10}\partial^{-1}(u_{2}u_{9}^{*}+u_{3}u_{8}^{*}+u_{6}u_{7}^{*})(\alpha-\beta)+\frac{1}{4}u_{10}\partial^{-1}(u_{2}^{*}u_{9}+u_{3}^{*}u_{8}+u_{6}^{*}u_{7})(\alpha-\beta)\\&+\frac{1}{4}u_{10}^{*}\partial^{-1}(u_{2}u_{9}+u_{3}u_{8}+u_{6}u_{7})(\alpha-\beta)+\frac{1}{4}u_{12}\partial^{-1}(u_{1}u_{6}+u_{2}u_{11}+u_{8}u_{9})(\alpha^{*}-\gamma^{*})\\&+\frac{1}{4}u_{12}\partial^{-1}(u_{1}u_{6}^{*}+u_{2}u_{11}^{*}+u_{8}u_{9}^{*})(\alpha-\gamma)+\frac{1}{4}u_{12}\partial^{-1}(u_{1}^{*}u_{6}+u_{2}^{*}u_{11}+u_{8}^{*}u_{9})(\alpha-\gamma)\\&+\frac{1}{4}u_{12}^{*}\partial^{-1}(u_{1}u_{6}+u_{2}u_{11}+u_{8}u_{9})(\alpha-\gamma),\\
		m_{2}^{*}=&\frac{1}{2}u_{3x}\beta^{*}+\frac{1}{2}u_{3x}^{*}\beta+\frac{1}{2}u_{7}\partial^{-1}(u_{2}u_{9}+u_{3}u_{8}+u_{6}u_{7})(\alpha^{*}-\beta^{*})+\frac{1}{2}u_{7}\partial^{-1}(u_{2}u_{9}^{*}+u_{3}u_{8}^{*}+u_{6}u_{7}^{*})(\alpha-\beta)\\&+\frac{1}{2}u_{7}\partial^{-1}(u_{2}^{*}u_{9}+u_{3}^{*}u_{8}+u_{6}^{*}u_{7})(\alpha-\beta)+\frac{1}{2}u_{7}^{*}\partial^{-1}(u_{2}u_{9}+u_{3}u_{8}+u_{6}u_{7})(\alpha-\beta)\\&+\frac{1}{2}u_{9}\partial^{-1}(u_{2}u_{7}+u_{3}u_{10}+u_{9}u_{12})(\gamma^{*}-\beta^{*})+\frac{1}{2}u_{9}\partial^{-1}(u_{2}u_{7}^{*}+u_{3}u_{10}^{*}+u_{9}u_{12}^{*})(\gamma-\beta)\\&+\frac{1}{2}u_{9}\partial^{-1}(u_{2}^{*}u_{7}+u_{3}^{*}u_{10}+u_{9}^{*}u_{12})(\gamma-\beta)+\frac{1}{2}u_{9}^{*}\partial^{-1}(u_{2}u_{7}+u_{3}u_{10}+u_{9}u_{12})(\gamma-\beta),\\
		o_{2}^{*}=&-\frac{1}{2}u_{4x}\beta^{*}-\frac{1}{2}u_{4x}^{*}\beta+\frac{1}{2}u_{8}\partial^{-1}(u_{1}u_{10}+u_{4}u_{7}+u_{5}u_{8})(\beta^{*}-\alpha^{*})+\frac{1}{2}u_{8}\partial^{-1}(u_{1}u_{10}^{*}+u_{4}u_{7}^{*}+u_{5}u_{8}^{*})(\beta-\alpha)\\&+\frac{1}{2}u_{8}\partial^{-1}(u_{1}^{*}u_{10}+u_{4}^{*}u_{7}+u_{5}^{*}u_{8})(\beta-\alpha)+\frac{1}{2}u_{8}^{*}\partial^{-1}(u_{1}u_{10}+u_{4}u_{7}+u_{5}u_{8})(\beta-\alpha)\\&+\frac{1}{2}u_{10}\partial^{-1}(u_{1}u_{8}+u_{4}u_{9}+u_{10}u_{11})(\beta^{*}-\gamma^{*})+\frac{1}{2}u_{10}\partial^{-1}(u_{1}u_{8}^{*}+u_{4}u_{9}^{*}+u_{10}u_{11}^{*})(\beta-\gamma)\\&+\frac{1}{2}u_{10}\partial^{-1}(u_{1}^{*}u_{8}+u_{4}^{*}u_{9}+u_{10}^{*}u_{11})(\beta-\gamma)+\frac{1}{2}u_{10}^{*}\partial^{-1}(u_{1}u_{8}+u_{4}u_{9}+u_{10}u_{11})(\beta-\gamma),\\
		p_{2}^{*}=&\frac{1}{2}u_{5x}\alpha^{*}+\frac{1}{2}u_{5x}^{*}\alpha+\frac{1}{2}u_{1}\partial^{-1}(u_{1}u_{12}+u_{2}u_{5}+u_{7}u_{10})(\gamma^{*}-\alpha^{*})+\frac{1}{2}u_{1}\partial^{-1}(u_{1}u_{12}^{*}+u_{2}u_{5}^{*}+u_{7}u_{10}^{*})(\gamma-\alpha)\\&+\frac{1}{2}u_{1}\partial^{-1}(u_{1}^{*}u_{12}+u_{2}^{*}u_{5}+u_{7}^{*}u_{10})(\gamma-\alpha)+\frac{1}{2}u_{1}^{*}\partial^{-1}(u_{1}u_{12}+u_{2}u_{5}+u_{7}u_{10})(\gamma-\alpha)\\&+\frac{1}{2}u_{7}\partial^{-1}(u_{1}u_{10}+u_{4}u_{7}+u_{5}u_{8})(\beta^{*}-\alpha^{*})+\frac{1}{2}u_{7}\partial^{-1}(u_{1}u_{10}^{*}+u_{4}u_{7}^{*}+u_{5}u_{8}^{*})(\beta-\alpha)\\&+\frac{1}{2}u_{7}\partial^{-1}(u_{1}^{*}u_{10}+u_{4}^{*}u_{7}+u_{5}^{*}u_{8})(\beta-\alpha)+\frac{1}{2}u_{7}^{*}\partial^{-1}(u_{1}u_{10}+u_{4}u_{7}+u_{5}u_{8})(\beta-\alpha),\\
		q_{2}^{*}=&-\frac{1}{2}u_{6x}\alpha^{*}-\frac{1}{2}u_{6x}^{*}\alpha+\frac{1}{2}u_{2}\partial^{-1}(u_{1}u_{6}+u_{2}u_{11}+u_{8}u_{9})(\alpha^{*}-\gamma^{*})+\frac{1}{2}u_{2}\partial^{-1}(u_{1}u_{6}^{*}+u_{2}u_{11}^{*}+u_{8}u_{9}^{*})(\alpha-\gamma)\\&+\frac{1}{2}u_{2}\partial^{-1}(u_{1}^{*}u_{6}+u_{2}^{*}u_{11}+u_{8}^{*}u_{9})(\alpha-\gamma)+\frac{1}{2}u_{2}^{*}\partial^{-1}(u_{1}u_{6}+u_{2}u_{11}+u_{8}u_{9})(\alpha-\gamma)\\&+\frac{1}{2}u_{8}\partial^{-1}(u_{2}u_{9}+u_{3}u_{8}+u_{6}u_{7})(\alpha^{*}-\beta^{*})+\frac{1}{2}u_{8}\partial^{-1}(u_{2}u_{9}^{*}+u_{3}u_{8}^{*}+u_{6}u_{7}^{*})(\alpha-\beta)\\&+\frac{1}{2}u_{8}\partial^{-1}(u_{2}^{*}u_{9}+u_{3}^{*}u_{8}+u_{6}^{*}u_{7})(\alpha-\beta)+\frac{1}{2}u_{8}^{*}\partial^{-1}(u_{2}u_{9}+u_{3}u_{8}+u_{6}u_{7})(\alpha-\beta),\\
		r_{2}^{*}=&\frac{1}{4}u_{7x}(\alpha^{*}+\beta^{*})+\frac{1}{4}u_{7x}^{*}(\alpha+\beta)+\frac{1}{4}u_{1}\partial^{-1}(u_{2}u_{7}+u_{3}u_{10}+u_{9}u_{12})(\gamma^{*}-\beta^{*})\\&+\frac{1}{4}u_{1}\partial^{-1}(u_{2}u_{7}^{*}+u_{3}u_{10}^{*}+u_{9}u_{12}^{*})(\gamma-\beta)+\frac{1}{4}u_{1}\partial^{-1}(u_{2}^{*}u_{7}+u_{3}^{*}u_{10}+u_{9}^{*}u_{12})(\gamma-\beta)\\&+\frac{1}{4}u_{1}^{*}\partial^{-1}(u_{2}u_{7}+u_{3}u_{10}+u_{9}u_{12})(\gamma-\beta)+\frac{1}{4}u_{3}\partial^{-1}(u_{1}u_{10}+u_{4}u_{7}+u_{5}u_{8})(\beta^{*}-\alpha^{*})\\&+\frac{1}{4}u_{3}\partial^{-1}(u_{1}u_{10}^{*}+u_{4}u_{7}^{*}+u_{5}u_{8}^{*})(\beta-\alpha)+\frac{1}{4}u_{3}\partial^{-1}(u_{1}^{*}u_{10}+u_{4}^{*}u_{7}+u_{5}^{*}u_{8})(\beta-\alpha)\\&+\frac{1}{4}u_{3}^{*}\partial^{-1}(u_{1}u_{10}+u_{4}u_{7}+u_{5}u_{8})(\beta-\alpha)+\frac{1}{4}u_{5}\partial^{-1}(u_{2}u_{9}+u_{3}u_{8}+u_{6}u_{7})(\alpha^{*}-\beta^{*})\\&+\frac{1}{4}u_{5}\partial^{-1}(u_{2}u_{9}^{*}+u_{3}u_{8}^{*}+u_{6}u_{7}^{*})(\alpha-\beta)+\frac{1}{4}u_{5}\partial^{-1}(u_{2}^{*}u_{9}+u_{3}^{*}u_{8}+u_{6}^{*}u_{7})(\alpha-\beta)\\&+\frac{1}{4}u_{5}^{*}\partial^{-1}(u_{2}u_{9}+u_{3}u_{8}+u_{6}u_{7})(\alpha-\beta)+\frac{1}{4}u_{9}\partial^{-1}(u_{1}u_{12}+u_{2}u_{5}+u_{7}u_{10})(\gamma^{*}-\alpha^{*})\\&+\frac{1}{4}u_{9}\partial^{-1}(u_{1}u_{12}^{*}+u_{2}u_{5}^{*}+u_{7}u_{10}^{*})(\gamma-\alpha)+\frac{1}{4}u_{9}\partial^{-1}(u_{1}^{*}u_{12}+u_{2}^{*}u_{5}+u_{7}^{*}u_{10})(\gamma-\alpha)\\&+\frac{1}{4}u_{9}^{*}\partial^{-1}(u_{1}u_{12}+u_{2}u_{5}+u_{7}u_{10})(\gamma-\alpha),\\
		s_{2}^{*}=&-\frac{1}{4}u_{8x}(\alpha^{*}+\beta^{*})-\frac{1}{4}u_{8x}^{*}(\alpha+\beta)+\frac{1}{4}u_{2}\partial^{-1}(u_{1}u_{8}+u_{4}u_{9}+u_{10}u_{11})(\beta^{*}-\gamma^{*})\\&+\frac{1}{4}u_{2}\partial^{-1}(u_{1}u_{8}^{*}+u_{4}u_{9}^{*}+u_{10}u_{11}^{*})(\beta-\gamma)+\frac{1}{4}u_{2}\partial^{-1}(u_{1}^{*}u_{8}+u_{4}^{*}u_{9}+u_{10}^{*}u_{11})(\beta-\gamma)\\&+\frac{1}{4}u_{2}^{*}\partial^{-1}(u_{1}u_{8}+u_{4}u_{9}+u_{10}u_{11})(\beta-\gamma)+\frac{1}{4}u_{4}\partial^{-1}(u_{2}u_{9}+u_{3}u_{8}+u_{6}u_{7})(\alpha^{*}-\beta^{*})\\&+\frac{1}{4}u_{4}\partial^{-1}(u_{2}u_{9}^{*}+u_{3}u_{8}^{*}+u_{6}u_{7}^{*})(\alpha-\beta)+\frac{1}{4}u_{4}\partial^{-1}(u_{2}^{*}u_{9}+u_{3}^{*}u_{8}+u_{6}^{*}u_{7})(\alpha-\beta)\\&+\frac{1}{4}u_{4}^{*}\partial^{-1}(u_{2}u_{9}+u_{3}u_{8}+u_{6}u_{7})(\alpha-\beta)+\frac{1}{4}u_{6}\partial^{-1}(u_{1}u_{10}+u_{4}u_{7}+u_{5}u_{8})(\beta^{*}-\alpha^{*})\\&+\frac{1}{4}u_{6}\partial^{-1}(u_{1}u_{10}^{*}+u_{4}u_{7}^{*}+u_{5}u_{8}^{*})(\beta-\alpha)+\frac{1}{4}u_{6}\partial^{-1}(u_{1}^{*}u_{10}+u_{4}^{*}u_{7}+u_{5}^{*}u_{8})(\beta-\alpha)\\&+\frac{1}{4}u_{6}^{*}\partial^{-1}(u_{1}u_{10}+u_{4}u_{7}+u_{5}u_{8})(\beta-\alpha)+\frac{1}{4}u_{10}\partial^{-1}(u_{1}u_{6}+u_{2}u_{11}+u_{8}u_{9})(\alpha^{*}-\gamma^{*})\\&+\frac{1}{4}u_{10}\partial^{-1}(u_{1}u_{6}^{*}+u_{2}u_{11}^{*}+u_{8}u_{9}^{*})(\alpha-\gamma)+\frac{1}{4}u_{10}\partial^{-1}(u_{1}^{*}u_{6}+u_{2}^{*}u_{11}+u_{8}^{*}u_{9})(\alpha-\gamma)\\&+\frac{1}{4}u_{10}^{*}\partial^{-1}(u_{1}u_{6}+u_{2}u_{11}+u_{8}u_{9})(\alpha-\gamma),\\
		t_{2}^{*}=&\frac{1}{4}u_{9x}(\beta^{*}+\gamma^{*})+\frac{1}{4}u_{9x}^{*}(\beta+\gamma)+\frac{1}{4}u_{1}\partial^{-1}(u_{2}u_{9}+u_{3}u_{8}+u_{6}u_{7})(\alpha^{*}-\beta^{*})\\&+\frac{1}{4}u_{1}\partial^{-1}(u_{2}u_{9}^{*}+u_{3}u_{8}^{*}+u_{6}u_{7}^{*})(\alpha-\beta)+\frac{1}{4}u_{1}\partial^{-1}(u_{2}^{*}u_{9}+u_{3}^{*}u_{8}+u_{6}^{*}u_{7})(\alpha-\beta)\\&+\frac{1}{4}u_{1}^{*}\partial^{-1}(u_{2}u_{9}+u_{3}u_{8}+u_{6}u_{7})(\alpha-\beta)+\frac{1}{4}u_{3}\partial^{-1}(u_{1}u_{8}+u_{4}u_{9}+u_{10}u_{11})(\beta^{*}-\gamma^{*})\\&+\frac{1}{4}u_{3}\partial^{-1}(u_{1}u_{8}^{*}+u_{4}u_{9}^{*}+u_{10}u_{11}^{*})(\beta-\gamma)+\frac{1}{4}u_{3}\partial^{-1}(u_{1}^{*}u_{8}+u_{4}^{*}u_{9}+u_{10}^{*}u_{11})(\beta-\gamma)\\&+\frac{1}{4}u_{3}^{*}\partial^{-1}(u_{1}u_{8}+u_{4}u_{9}+u_{10}u_{11})(\beta-\gamma)+\frac{1}{4}u_{7}\partial^{-1}(u_{1}u_{6}+u_{2}u_{11}+u_{8}u_{9})(\alpha^{*}-\gamma^{*})\\&+\frac{1}{4}u_{7}\partial^{-1}(u_{1}u_{6}^{*}+u_{2}u_{11}^{*}+u_{8}u_{9}^{*})(\alpha-\gamma)+\frac{1}{4}u_{7}\partial^{-1}(u_{1}^{*}u_{6}+u_{2}^{*}u_{11}+u_{8}^{*}u_{9})(\alpha-\gamma)\\&+\frac{1}{4}u_{7}^{*}\partial^{-1}(u_{1}u_{6}+u_{2}u_{11}+u_{8}u_{9})(\alpha-\gamma)+\frac{1}{4}u_{11}\partial^{-1}(u_{2}u_{7}+u_{3}u_{10}+u_{9}u_{12})(\gamma^{*}-\beta^{*})\\&+\frac{1}{4}u_{11}\partial^{-1}(u_{2}u_{7}^{*}+u_{3}u_{10}^{*}+u_{9}u_{12}^{*})(\gamma-\beta)+\frac{1}{4}u_{11}\partial^{-1}(u_{2}^{*}u_{7}+u_{3}^{*}u_{10}+u_{9}^{*}u_{12})(\gamma-\beta)\\&+\frac{1}{4}u_{11}^{*}\partial^{-1}(u_{2}u_{7}+u_{3}u_{10}+u_{9}u_{12})(\gamma-\beta),\\
		u_{2}^{*}=&-\frac{1}{4}u_{10x}(\beta^{*}+\gamma^{*})-\frac{1}{4}u_{10x}^{*}(\beta+\gamma)+\frac{1}{4}u_{2}\partial^{-1}(u_{1}u_{10}+u_{4}u_{7}+u_{5}u_{8})(\beta^{*}-\alpha^{*})\\&+\frac{1}{4}u_{2}\partial^{-1}(u_{1}u_{10}^{*}+u_{4}u_{7}^{*}+u_{5}u_{8}^{*})(\beta-\alpha)+\frac{1}{4}u_{2}\partial^{-1}(u_{1}^{*}u_{10}+u_{4}^{*}u_{7}+u_{5}^{*}u_{8})(\beta-\alpha)\\&+\frac{1}{4}u_{2}^{*}\partial^{-1}(u_{1}u_{10}+u_{4}u_{7}+u_{5}u_{8})(\beta-\alpha)+\frac{1}{4}u_{4}\partial^{-1}(u_{2}u_{7}+u_{3}u_{10}+u_{9}u_{12})(\gamma^{*}-\beta^{*})\\&+\frac{1}{4}u_{4}\partial^{-1}(u_{2}u_{7}^{*}+u_{3}u_{10}^{*}+u_{9}u_{12}^{*})(\gamma-\beta)+\frac{1}{4}u_{4}\partial^{-1}(u_{2}^{*}u_{7}+u_{3}^{*}u_{10}+u_{9}^{*}u_{12})(\gamma-\beta)\\&+\frac{1}{4}u_{4}^{*}\partial^{-1}(u_{2}u_{7}+u_{3}u_{10}+u_{9}u_{12})(\gamma-\beta)+\frac{1}{4}u_{8}\partial^{-1}(u_{1}u_{12}+u_{2}u_{5}+u_{7}u_{10})(\gamma^{*}-\alpha^{*})\\&+\frac{1}{4}u_{8}\partial^{-1}(u_{1}u_{12}^{*}+u_{2}u_{5}^{*}+u_{7}u_{10}^{*})(\gamma-\alpha)+\frac{1}{4}u_{8}\partial^{-1}(u_{1}^{*}u_{12}+u_{2}^{*}u_{5}+u_{7}^{*}u_{10})(\gamma-\alpha)\\&+\frac{1}{4}u_{8}^{*}\partial^{-1}(u_{1}u_{12}+u_{2}u_{5}+u_{7}u_{10})(\gamma-\alpha)+\frac{1}{4}u_{12}\partial^{-1}(u_{1}u_{8}+u_{4}u_{9}+u_{10}u_{11})(\beta^{*}-\gamma^{*})\\&+\frac{1}{4}u_{12}\partial^{-1}(u_{1}u_{8}^{*}+u_{4}u_{9}^{*}+u_{10}u_{11}^{*})(\beta-\gamma)+\frac{1}{4}u_{12}\partial^{-1}(u_{1}^{*}u_{8}+u_{4}^{*}u_{9}+u_{10}^{*}u_{11})(\beta-\gamma)\\&+\frac{1}{4}u_{12}^{*}\partial^{-1}(u_{1}u_{8}+u_{4}u_{9}+u_{10}u_{11})(\beta-\gamma),\\
		v_{2}^{*}=&\frac{1}{2}u_{11x}\gamma^{*}+\frac{1}{2}u_{11x}^{*}\gamma+\frac{1}{2}u_{1}\partial^{-1}(u_{1}u_{6}+u_{2}u_{11}+u_{8}u_{9})(\alpha^{*}-\gamma^{*})+\frac{1}{2}u_{1}\partial^{-1}(u_{1}u_{6}^{*}+u_{2}u_{11}^{*}+u_{8}u_{9}^{*})(\alpha-\gamma)\\&+\frac{1}{2}u_{1}\partial^{-1}(u_{1}^{*}u_{6}+u_{2}^{*}u_{11}+u_{8}^{*}u_{9})(\alpha-\gamma)+\frac{1}{2}u_{1}^{*}\partial^{-1}(u_{1}u_{6}+u_{2}u_{11}+u_{8}u_{9})(\alpha-\gamma)\\&+\frac{1}{2}u_{9}\partial^{-1}(u_{1}u_{8}+u_{4}u_{9}+u_{10}u_{11})(\beta^{*}-\gamma^{*})+\frac{1}{2}u_{9}\partial^{-1}(u_{1}u_{8}^{*}+u_{4}u_{9}^{*}+u_{10}u_{11}^{*})(\beta-\gamma)\\&+\frac{1}{2}u_{9}\partial^{-1}(u_{1}^{*}u_{8}+u_{4}^{*}u_{9}+u_{10}^{*}u_{11})(\beta-\gamma)+\frac{1}{2}u_{9}^{*}\partial^{-1}(u_{1}u_{8}+u_{4}u_{9}+u_{10}u_{11})(\beta-\gamma),\\
		w_{2}^{*}=&-\frac{1}{2}u_{12x}\gamma^{*}-\frac{1}{2}u_{12x}^{*}\gamma+\frac{1}{2}u_{2}\partial^{-1}(u_{1}u_{12}+u_{2}u_{5}+u_{7}u_{10})(\gamma^{*}-\alpha^{*})+\frac{1}{2}u_{2}\partial^{-1}(u_{1}u_{12}^{*}+u_{2}u_{5}^{*}+u_{7}u_{10}^{*})(\gamma-\alpha)\\&+\frac{1}{2}u_{2}\partial^{-1}(u_{1}^{*}u_{12}+u_{2}^{*}u_{5}+u_{7}^{*}u_{10})(\gamma-\alpha)+\frac{1}{2}u_{2}^{*}\partial^{-1}(u_{1}u_{12}+u_{2}u_{5}+u_{7}u_{10})(\gamma-\alpha)\\&+\frac{1}{2}u_{10}\partial^{-1}(u_{2}u_{7}+u_{3}u_{10}+u_{9}u_{12})(\gamma^{*}-\beta^{*})+\frac{1}{2}u_{10}\partial^{-1}(u_{2}u_{7}^{*}+u_{3}u_{10}^{*}+u_{9}u_{12}^{*})(\gamma-\beta)\\&+\frac{1}{2}u_{10}\partial^{-1}(u_{2}^{*}u_{7}+u_{3}^{*}u_{10}+u_{9}^{*}u_{12})(\gamma-\beta)+\frac{1}{2}u_{10}^{*}\partial^{-1}(u_{2}u_{7}+u_{3}u_{10}+u_{9}u_{12})(\gamma-\beta),\\
		\dots&\dots
	\end{align*}
	From the zero curvature equation $U_{3,t}-V_{3,+x}^{n}+[U_{3},V_{3,+}^{n}]=0$, we can obtain the following Lax integrable hierarchy
	\begin{align}\label{eq:3.2}
		\hat{u}_{t}=\left(\begin{array}{c}
			u\\
			u^{*}
		\end{array}\right)_{t}=
		\left(\begin{array}{cc}
			J_{1}&0\\
			0&J_1
		\end{array}\right)\left(\begin{array}{c}
			P_{1,n+1}\\
			P_{1,n+1}^{*}
		\end{array}\right)=JP_{n+1},
	\end{align}
	where $u_{t},\ J_{1},\ P_{1,n+1}$ be defined as (\ref{eq:2.5}), and
	\begin{align*}
		u_{t}^{*}=\left(
		\begin{array}{c}
			2k_{n+1}^{*}\\
			-2l_{n+1}^{*}\\
			2m_{n+1}^{*}\\
			-2o_{n+1}^{*}\\
			2p_{n+1}^{*}\\
			-2q_{n+1}^{*}\\
			2r_{n+1}^{*}\\
			-2s_{n+1}^{*}\\
			2t_{n+1}^{*}\\
			-2u_{n+1}^{*}\\
			2v_{n+1}^{*}\\
			-2w_{n+1}^{*}\\
		\end{array}\right),
		P_{1,n+1}^{*}=\left(
		\begin{array}{c}
			2l_{n+1}^{*}\\
			2k_{n+1}^{*}\\
			o_{n+1}^{*}\\
			m_{n+1}^{*}\\
			q_{n+1}^{*}\\
			p_{n+1}^{*}\\
			2s_{n+1}^{*}\\
			2r_{n+1}^{*}\\
			2u_{n+1}^{*}\\
			2t_{n+1}^{*}\\
			w_{n+1}^{*}\\
			v_{n+1}^{*}\\
		\end{array}\right).
	\end{align*}
	Thus, employing the Kronecker product, we have derived the integrable coupling system for the soliton hierarchy (\ref{eq:2.5}).
	When n=1, the hierarchy (\ref{eq:3.2}) reduces to the first system (\ref{eq:2.6}) and
	\begin{align*}
		u_{1t}*=&\frac{1}{2}u_{1x}(\alpha^{*}+\gamma^{*})+\frac{1}{2}u_{1x}^{*}(\alpha+\gamma)+\frac{1}{2}u_{5}\partial^{-1}(u_{1}u_{6}+u_{2}u_{11}+u_{8}u_{9})(\alpha^{*}-\gamma^{*})\\
		&+\frac{1}{2}u_{5}\partial^{-1}(u_{1}u_{6}^{*}+u_{2}u_{11}^{*}+u_{8}u_{9}^{*})(\alpha-\gamma)+\frac{1}{2}u_{5}\partial^{-1}(u_{1}^{*}u_{6}+u_{2}^{*}u_{11}+u_{8}^{*}u_{9})(\alpha-\gamma)\\
		&+\frac{1}{2}u_{5}^{*}\partial^{-1}(u_{1}u_{6}+u_{2}u_{11}+u_{8}u_{9})(\alpha-\gamma)+\frac{1}{2}u_{7}\partial^{-1}(u_{1}u_{8}+u_{4}u_{9}+u_{10}u_{11})(\beta^{*}-\gamma^{*})\\
		&+\frac{1}{2}u_{7}\partial^{-1}(u_{1}u_{8}^{*}+u_{4}u_{9}^{*}+u_{10}u_{11}^{*})(\beta-\gamma)+\frac{1}{2}u_{7}\partial^{-1}(u_{1}^{*}u_{8}+u_{4}^{*}u_{9}+u_{10}^{*}u_{11})(\beta-\gamma)\\
		&+\frac{1}{2}u_{7}^{*}\partial^{-1}(u_{1}u_{8}+u_{4}u_{9}+u_{10}u_{11})(\beta-\gamma)+\frac{1}{2}u_{9}\partial^{-1}(u_{1}u_{10}+u_{4}u_{7}+u_{5}u_{8})(\beta^{*}-\alpha^{*})\\
		&+\frac{1}{2}u_{9}\partial^{-1}(u_{1}u_{10}^{*}+u_{4}u_{7}^{*}+u_{5}u_{8}^{*})(\beta-\alpha)+\frac{1}{2}u_{9}\partial^{-1}(u_{1}^{*}u_{10}+u_{4}^{*}u_{7}+u_{5}^{*}u_{8})(\beta-\alpha)\\
		&+\frac{1}{2}u_{9}^{*}\partial^{-1}(u_{1}u_{10}+u_{4}u_{7}+u_{5}u_{8})(\beta-\alpha)+\frac{1}{2}u_{11}\partial^{-1}(u_{1}u_{12}+u_{2}u_{5}+u_{7}u_{10})(\gamma^{*}-\alpha^{*})\\
		&+\frac{1}{2}u_{11}\partial^{-1}(u_{1}u_{12}^{*}+u_{2}u_{5}^{*}+u_{7}u_{10}^{*})(\gamma-\alpha)+\frac{1}{2}u_{11}\partial^{-1}(u_{1}^{*}u_{12}+u_{2}^{*}u_{5}+u_{7}^{*}u_{10})(\gamma-\alpha)\\
		&+\frac{1}{2}u_{11}^{*}\partial^{-1}(u_{1}u_{12}+u_{2}u_{5}+u_{7}u_{10})(\gamma-\alpha),\\ u_{2t}*=&\frac{1}{2}u_{2x}(\alpha^{*}+\gamma^{*})+\frac{1}{2}u_{2x}^{*}(\alpha+\gamma)-\frac{1}{2}u_{6}\partial^{-1}(u_{1}u_{12}+u_{2}u_{5}+u_{7}u_{10})(\gamma^{*}-\alpha^{*})\\
		&-\frac{1}{2}u_{6}\partial^{-1}(u_{1}u_{12}^{*}+u_{2}u_{5}^{*}+u_{7}u_{10}^{*})(\gamma-\alpha)-\frac{1}{2}u_{6}\partial^{-1}(u_{1}^{*}u_{12}+u_{2}^{*}u_{5}+u_{7}^{*}u_{10})(\gamma-\alpha)\\
		&-\frac{1}{2}u_{6}^{*}\partial^{-1}(u_{1}u_{12}+u_{2}u_{5}+u_{7}u_{10})(\gamma-\alpha)-\frac{1}{2}u_{8}\partial^{-1}(u_{2}u_{7}+u_{3}u_{10}+u_{9}u_{12})(\gamma^{*}-\beta^{*})\\
		&-\frac{1}{2}u_{8}\partial^{-1}(u_{2}u_{7}^{*}+u_{3}u_{10}^{*}+u_{9}u_{12}^{*})(\gamma-\beta)-\frac{1}{2}u_{8}\partial^{-1}(u_{2}^{*}u_{7}+u_{3}^{*}u_{10}+u_{9}^{*}u_{12})(\gamma-\beta)\\&-\frac{1}{2}u_{8}^{*}\partial^{-1}(u_{2}u_{7}+u_{3}u_{10}+u_{9}u_{12})(\gamma-\beta)-\frac{1}{2}u_{10}\partial^{-1}(u_{2}u_{9}+u_{3}u_{8}+u_{6}u_{7})(\alpha^{*}-\beta^{*})\\
		&-\frac{1}{2}u_{10}\partial^{-1}(u_{2}u_{9}^{*}+u_{3}u_{8}^{*}+u_{6}u_{7}^{*})(\alpha-\beta)-\frac{1}{2}u_{10}\partial^{-1}(u_{2}^{*}u_{9}+u_{3}^{*}u_{8}+u_{6}^{*}u_{7})(\alpha-\beta)\\&-\frac{1}{2}u_{10}^{*}\partial^{-1}(u_{2}u_{9}+u_{3}u_{8}+u_{6}u_{7})(\alpha-\beta)-\frac{1}{2}u_{12}\partial^{-1}(u_{1}u_{6}+u_{2}u_{11}+u_{8}u_{9})(\alpha^{*}-\gamma^{*})\\&-\frac{1}{2}u_{12}\partial^{-1}(u_{1}u_{6}^{*}+u_{2}u_{11}^{*}+u_{8}u_{9}^{*})(\alpha-\gamma)-\frac{1}{2}u_{12}\partial^{-1}(u_{1}^{*}u_{6}+u_{2}^{*}u_{11}+u_{8}^{*}u_{9})(\alpha-\gamma)\\&-\frac{1}{2}u_{12}^{*}\partial^{-1}(u_{1}u_{6}+u_{2}u_{11}+u_{8}u_{9})(\alpha-\gamma),\\	u_{3t}*=&u_{3x}\beta^{*}+u_{3x}^{*}\beta+u_{7}\partial^{-1}(u_{2}u_{9}+u_{3}u_{8}+u_{6}u_{7})(\alpha^{*}-\beta^{*})+u_{7}\partial^{-1}(u_{2}u_{9}^{*}+u_{3}u_{8}^{*}+u_{6}u_{7}^{*})(\alpha-\beta)\\&+u_{7}\partial^{-1}(u_{2}^{*}u_{9}+u_{3}^{*}u_{8}+u_{6}^{*}u_{7})(\alpha-\beta)+u_{7}^{*}\partial^{-1}(u_{2}u_{9}+u_{3}u_{8}+u_{6}u_{7})(\alpha-\beta)\\&+u_{9}\partial^{-1}(u_{2}u_{7}+u_{3}u_{10}+u_{9}u_{12})(\gamma^{*}-\beta^{*})+u_{9}\partial^{-1}(u_{2}u_{7}^{*}+u_{3}u_{10}^{*}+u_{9}u_{12}^{*})(\gamma-\beta)\\&+u_{9}\partial^{-1}(u_{2}^{*}u_{7}+u_{3}^{*}u_{10}+u_{9}^{*}u_{12})(\gamma-\beta)+u_{9}^{*}\partial^{-1}(u_{2}u_{7}+u_{3}u_{10}+u_{9}u_{12})(\gamma-\beta),\\
		u_{4t}*=&u_{4x}\beta^{*}+u_{4x}^{*}\beta-u_{8}\partial^{-1}(u_{1}u_{10}+u_{4}u_{7}+u_{5}u_{8})(\beta^{*}-\alpha^{*})-u_{8}\partial^{-1}(u_{1}u_{10}^{*}+u_{4}u_{7}^{*}+u_{5}u_{8}^{*})(\beta-\alpha)\\&-u_{8}\partial^{-1}(u_{1}^{*}u_{10}+u_{4}^{*}u_{7}+u_{5}^{*}u_{8})(\beta-\alpha)-u_{8}^{*}\partial^{-1}(u_{1}u_{10}+u_{4}u_{7}+u_{5}u_{8})(\beta-\alpha)\\&-u_{10}\partial^{-1}(u_{1}u_{8}+u_{4}u_{9}+u_{10}u_{11})(\beta^{*}-\gamma^{*})-u_{10}\partial^{-1}(u_{1}u_{8}^{*}+u_{4}u_{9}^{*}+u_{10}u_{11}^{*})(\beta-\gamma)\\&-u_{10}\partial^{-1}(u_{1}^{*}u_{8}+u_{4}^{*}u_{9}+u_{10}^{*}u_{11})(\beta-\gamma)-u_{10}^{*}\partial^{-1}(u_{1}u_{8}+u_{4}u_{9}+u_{10}u_{11})(\beta-\gamma),\\
		u_{5t}*=&u_{5x}\alpha^{*}+u_{5x}^{*}\alpha+u_{1}\partial^{-1}(u_{1}u_{12}+u_{2}u_{5}+u_{7}u_{10})(\gamma^{*}-\alpha^{*})+u_{1}\partial^{-1}(u_{1}u_{12}^{*}+u_{2}u_{5}^{*}+u_{7}u_{10}^{*})(\gamma-\alpha)\\&+u_{1}\partial^{-1}(u_{1}^{*}u_{12}+u_{2}^{*}u_{5}+u_{7}^{*}u_{10})(\gamma-\alpha)+u_{1}^{*}\partial^{-1}(u_{1}u_{12}+u_{2}u_{5}+u_{7}u_{10})(\gamma-\alpha)\\&+u_{7}\partial^{-1}(u_{1}u_{10}+u_{4}u_{7}+u_{5}u_{8})(\beta^{*}-\alpha^{*})+u_{7}\partial^{-1}(u_{1}u_{10}^{*}+u_{4}u_{7}^{*}+u_{5}u_{8}^{*})(\beta-\alpha)\\&+u_{7}\partial^{-1}(u_{1}^{*}u_{10}+u_{4}^{*}u_{7}+u_{5}^{*}u_{8})(\beta-\alpha)+u_{7}^{*}\partial^{-1}(u_{1}u_{10}+u_{4}u_{7}+u_{5}u_{8})(\beta-\alpha),\\
		u_{6t}*=&u_{6x}\alpha^{*}+u_{6x}^{*}\alpha-u_{2}\partial^{-1}(u_{1}u_{6}+u_{2}u_{11}+u_{8}u_{9})(\alpha^{*}-\gamma^{*})-u_{2}\partial^{-1}(u_{1}u_{6}^{*}+u_{2}u_{11}^{*}+u_{8}u_{9}^{*})(\alpha-\gamma)\\&-u_{2}\partial^{-1}(u_{1}^{*}u_{6}+u_{2}^{*}u_{11}+u_{8}^{*}u_{9})(\alpha-\gamma)-u_{2}^{*}\partial^{-1}(u_{1}u_{6}+u_{2}u_{11}+u_{8}u_{9})(\alpha-\gamma)\\&-u_{8}\partial^{-1}(u_{2}u_{9}+u_{3}u_{8}+u_{6}u_{7})(\alpha^{*}-\beta^{*})-u_{8}\partial^{-1}(u_{2}u_{9}^{*}+u_{3}u_{8}^{*}+u_{6}u_{7}^{*})(\alpha-\beta)\\&-u_{8}\partial^{-1}(u_{2}^{*}u_{9}+u_{3}^{*}u_{8}+u_{6}^{*}u_{7})(\alpha-\beta)-u_{8}^{*}\partial^{-1}(u_{2}u_{9}+u_{3}u_{8}+u_{6}u_{7})(\alpha-\beta),\\
		u_{7t}*=&\frac{1}{2}u_{7x}(\alpha^{*}+\beta^{*})+\frac{1}{2}u_{7x}^{*}(\alpha+\beta)+\frac{1}{2}u_{1}\partial^{-1}(u_{2}u_{7}+u_{3}u_{10}+u_{9}u_{12})(\gamma^{*}-\beta^{*})\\&+\frac{1}{2}u_{1}\partial^{-1}(u_{2}u_{7}^{*}+u_{3}u_{10}^{*}+u_{9}u_{12}^{*})(\gamma-\beta)+\frac{1}{2}u_{1}\partial^{-1}(u_{2}^{*}u_{7}+u_{3}^{*}u_{10}+u_{9}^{*}u_{12})(\gamma-\beta)\\&+\frac{1}{2}u_{1}^{*}\partial^{-1}(u_{2}u_{7}+u_{3}u_{10}+u_{9}u_{12})(\gamma-\beta)+\frac{1}{2}u_{3}\partial^{-1}(u_{1}u_{10}+u_{4}u_{7}+u_{5}u_{8})(\beta^{*}-\alpha^{*})\\&+\frac{1}{2}u_{3}\partial^{-1}(u_{1}u_{10}^{*}+u_{4}u_{7}^{*}+u_{5}u_{8}^{*})(\beta-\alpha)+\frac{1}{2}u_{3}\partial^{-1}(u_{1}^{*}u_{10}+u_{4}^{*}u_{7}+u_{5}^{*}u_{8})(\beta-\alpha)\\&+\frac{1}{2}u_{3}^{*}\partial^{-1}(u_{1}u_{10}+u_{4}u_{7}+u_{5}u_{8})(\beta-\alpha)+\frac{1}{2}u_{5}\partial^{-1}(u_{2}u_{9}+u_{3}u_{8}+u_{6}u_{7})(\alpha^{*}-\beta^{*})\\&+\frac{1}{2}u_{5}\partial^{-1}(u_{2}u_{9}^{*}+u_{3}u_{8}^{*}+u_{6}u_{7}^{*})(\alpha-\beta)+\frac{1}{2}u_{5}\partial^{-1}(u_{2}^{*}u_{9}+u_{3}^{*}u_{8}+u_{6}^{*}u_{7})(\alpha-\beta)\\&+\frac{1}{4}u_{5}^{*}\partial^{-1}(u_{2}u_{9}+u_{3}u_{8}+u_{6}u_{7})(\alpha-\beta)+\frac{1}{2}u_{9}\partial^{-1}(u_{1}u_{12}+u_{2}u_{5}+u_{7}u_{10})(\gamma^{*}-\alpha^{*})\\&+\frac{1}{2}u_{9}\partial^{-1}(u_{1}u_{12}^{*}+u_{2}u_{5}^{*}+u_{7}u_{10}^{*})(\gamma-\alpha)+\frac{1}{2}u_{9}\partial^{-1}(u_{1}^{*}u_{12}+u_{2}^{*}u_{5}+u_{7}^{*}u_{10})(\gamma-\alpha)\\&+\frac{1}{2}u_{9}^{*}\partial^{-1}(u_{1}u_{12}+u_{2}u_{5}+u_{7}u_{10})(\gamma-\alpha),\\
		u_{8t}*=&\frac{1}{2}u_{8x}(\alpha^{*}+\beta^{*})+\frac{1}{2}u_{8x}^{*}(\alpha+\beta)-\frac{1}{2}u_{2}\partial^{-1}(u_{1}u_{8}+u_{4}u_{9}+u_{10}u_{11})(\beta^{*}-\gamma^{*})\\&-\frac{1}{2}u_{2}\partial^{-1}(u_{1}u_{8}^{*}+u_{4}u_{9}^{*}+u_{10}u_{11}^{*})(\beta-\gamma)-\frac{1}{2}u_{2}\partial^{-1}(u_{1}^{*}u_{8}+u_{4}^{*}u_{9}+u_{10}^{*}u_{11})(\beta-\gamma)\\&-\frac{1}{2}u_{2}^{*}\partial^{-1}(u_{1}u_{8}+u_{4}u_{9}+u_{10}u_{11})(\beta-\gamma)-\frac{1}{2}u_{4}\partial^{-1}(u_{2}u_{9}+u_{3}u_{8}+u_{6}u_{7})(\alpha^{*}-\beta^{*})\\&-\frac{1}{2}u_{4}\partial^{-1}(u_{2}u_{9}^{*}+u_{3}u_{8}^{*}+u_{6}u_{7}^{*})(\alpha-\beta)-\frac{1}{2}u_{4}\partial^{-1}(u_{2}^{*}u_{9}+u_{3}^{*}u_{8}+u_{6}^{*}u_{7})(\alpha-\beta)\\&-\frac{1}{2}u_{4}^{*}\partial^{-1}(u_{2}u_{9}+u_{3}u_{8}+u_{6}u_{7})(\alpha-\beta)-\frac{1}{2}u_{6}\partial^{-1}(u_{1}u_{10}+u_{4}u_{7}+u_{5}u_{8})(\beta^{*}-\alpha^{*})\\&-\frac{1}{2}u_{6}\partial^{-1}(u_{1}u_{10}^{*}+u_{4}u_{7}^{*}+u_{5}u_{8}^{*})(\beta-\alpha)-\frac{1}{2}u_{6}\partial^{-1}(u_{1}^{*}u_{10}+u_{4}^{*}u_{7}+u_{5}^{*}u_{8})(\beta-\alpha)\\&-\frac{1}{2}u_{6}^{*}\partial^{-1}(u_{1}u_{10}+u_{4}u_{7}+u_{5}u_{8})(\beta-\alpha)-\frac{1}{2}u_{10}\partial^{-1}(u_{1}u_{6}+u_{2}u_{11}+u_{8}u_{9})(\alpha^{*}-\gamma^{*})\\&-\frac{1}{2}u_{10}\partial^{-1}(u_{1}u_{6}^{*}+u_{2}u_{11}^{*}+u_{8}u_{9}^{*})(\alpha-\gamma)-\frac{1}{2}u_{10}\partial^{-1}(u_{1}^{*}u_{6}+u_{2}^{*}u_{11}+u_{8}^{*}u_{9})(\alpha-\gamma)\\&-\frac{1}{2}u_{10}^{*}\partial^{-1}(u_{1}u_{6}+u_{2}u_{11}+u_{8}u_{9})(\alpha-\gamma),\\
		u_{9t}*=&\frac{1}{2}u_{9x}(\beta^{*}+\gamma^{*})+\frac{1}{2}u_{9x}^{*}(\beta+\gamma)+\frac{1}{2}u_{1}\partial^{-1}(u_{2}u_{9}+u_{3}u_{8}+u_{6}u_{7})(\alpha^{*}-\beta^{*})\\&+\frac{1}{2}u_{1}\partial^{-1}(u_{2}u_{9}^{*}+u_{3}u_{8}^{*}+u_{6}u_{7}^{*})(\alpha-\beta)+\frac{1}{2}u_{1}\partial^{-1}(u_{2}^{*}u_{9}+u_{3}^{*}u_{8}+u_{6}^{*}u_{7})(\alpha-\beta)\\&+\frac{1}{2}u_{1}^{*}\partial^{-1}(u_{2}u_{9}+u_{3}u_{8}+u_{6}u_{7})(\alpha-\beta)+\frac{1}{2}u_{3}\partial^{-1}(u_{1}u_{8}+u_{4}u_{9}+u_{10}u_{11})(\beta^{*}-\gamma^{*})\\&+\frac{1}{2}u_{3}\partial^{-1}(u_{1}u_{8}^{*}+u_{4}u_{9}^{*}+u_{10}u_{11}^{*})(\beta-\gamma)+\frac{1}{2}u_{3}\partial^{-1}(u_{1}^{*}u_{8}+u_{4}^{*}u_{9}+u_{10}^{*}u_{11})(\beta-\gamma)\\&+\frac{1}{2}u_{3}^{*}\partial^{-1}(u_{1}u_{8}+u_{4}u_{9}+u_{10}u_{11})(\beta-\gamma)+\frac{1}{2}u_{7}\partial^{-1}(u_{1}u_{6}+u_{2}u_{11}+u_{8}u_{9})(\alpha^{*}-\gamma^{*})\\&+\frac{1}{2}u_{7}\partial^{-1}(u_{1}u_{6}^{*}+u_{2}u_{11}^{*}+u_{8}u_{9}^{*})(\alpha-\gamma)+\frac{1}{2}u_{7}\partial^{-1}(u_{1}^{*}u_{6}+u_{2}^{*}u_{11}+u_{8}^{*}u_{9})(\alpha-\gamma)\\&+\frac{1}{2}u_{7}^{*}\partial^{-1}(u_{1}u_{6}+u_{2}u_{11}+u_{8}u_{9})(\alpha-\gamma)+\frac{1}{2}u_{11}\partial^{-1}(u_{2}u_{7}+u_{3}u_{10}+u_{9}u_{12})(\gamma^{*}-\beta^{*})\\&+\frac{1}{2}u_{11}\partial^{-1}(u_{2}u_{7}^{*}+u_{3}u_{10}^{*}+u_{9}u_{12}^{*})(\gamma-\beta)+\frac{1}{2}u_{11}\partial^{-1}(u_{2}^{*}u_{7}+u_{3}^{*}u_{10}+u_{9}^{*}u_{12})(\gamma-\beta)\\&+\frac{1}{2}u_{11}^{*}\partial^{-1}(u_{2}u_{7}+u_{3}u_{10}+u_{9}u_{12})(\gamma-\beta),\\
		u_{10t}*=&\frac{1}{2}u_{10x}(\beta^{*}+\gamma^{*})+\frac{1}{2}u_{10x}^{*}(\beta+\gamma)-\frac{1}{2}u_{2}\partial^{-1}(u_{1}u_{10}+u_{4}u_{7}+u_{5}u_{8})(\beta^{*}-\alpha^{*})\\&-\frac{1}{2}u_{2}\partial^{-1}(u_{1}u_{10}^{*}+u_{4}u_{7}^{*}+u_{5}u_{8}^{*})(\beta-\alpha)-\frac{1}{2}u_{2}\partial^{-1}(u_{1}^{*}u_{10}+u_{4}^{*}u_{7}+u_{5}^{*}u_{8})(\beta-\alpha)\\&-\frac{1}{2}u_{2}^{*}\partial^{-1}(u_{1}u_{10}+u_{4}u_{7}+u_{5}u_{8})(\beta-\alpha)-\frac{1}{2}u_{4}\partial^{-1}(u_{2}u_{7}+u_{3}u_{10}+u_{9}u_{12})(\gamma^{*}-\beta^{*})\\&-\frac{1}{2}u_{4}\partial^{-1}(u_{2}u_{7}^{*}+u_{3}u_{10}^{*}+u_{9}u_{12}^{*})(\gamma-\beta)-\frac{1}{2}u_{4}\partial^{-1}(u_{2}^{*}u_{7}+u_{3}^{*}u_{10}+u_{9}^{*}u_{12})(\gamma-\beta)\\
		&-\frac{1}{2}u_{4}^{*}\partial^{-1}(u_{2}u_{7}+u_{3}u_{10}+u_{9}u_{12})(\gamma-\beta)-\frac{1}{2}u_{8}\partial^{-1}(u_{1}u_{12}+u_{2}u_{5}+u_{7}u_{10})(\gamma^{*}-\alpha^{*})\\&-\frac{1}{2}u_{8}\partial^{-1}(u_{1}u_{12}^{*}+u_{2}u_{5}^{*}+u_{7}u_{10}^{*})(\gamma-\alpha)-\frac{1}{2}u_{8}\partial^{-1}(u_{1}^{*}u_{12}+u_{2}^{*}u_{5}+u_{7}^{*}u_{10})(\gamma-\alpha)\\&-\frac{1}{2}u_{8}^{*}\partial^{-1}(u_{1}u_{12}+u_{2}u_{5}+u_{7}u_{10})(\gamma-\alpha)-\frac{1}{2}u_{12}\partial^{-1}(u_{1}u_{8}+u_{4}u_{9}+u_{10}u_{11})(\beta^{*}-\gamma^{*})\\&-\frac{1}{2}u_{12}\partial^{-1}(u_{1}u_{8}^{*}+u_{4}u_{9}^{*}+u_{10}u_{11}^{*})(\beta-\gamma)-\frac{1}{2}u_{12}\partial^{-1}(u_{1}^{*}u_{8}+u_{4}^{*}u_{9}+u_{10}^{*}u_{11})(\beta-\gamma)\\&-\frac{1}{2}u_{12}^{*}\partial^{-1}(u_{1}u_{8}+u_{4}u_{9}+u_{10}u_{11})(\beta-\gamma),\\
		u_{11t}*=&u_{11x}\gamma^{*}+u_{11x}^{*}\gamma+u_{1}\partial^{-1}(u_{1}u_{6}+u_{2}u_{11}+u_{8}u_{9})(\alpha^{*}-\gamma^{*})+u_{1}\partial^{-1}(u_{1}u_{6}^{*}+u_{2}u_{11}^{*}+u_{8}u_{9}^{*})(\alpha-\gamma)\\&+u_{1}\partial^{-1}(u_{1}^{*}u_{6}+u_{2}^{*}u_{11}+u_{8}^{*}u_{9})(\alpha-\gamma)+u_{1}^{*}\partial^{-1}(u_{1}u_{6}+u_{2}u_{11}+u_{8}u_{9})(\alpha-\gamma)\\&+u_{9}\partial^{-1}(u_{1}u_{8}+u_{4}u_{9}+u_{10}u_{11})(\beta^{*}-\gamma^{*})+u_{9}\partial^{-1}(u_{1}u_{8}^{*}+u_{4}u_{9}^{*}+u_{10}u_{11}^{*})(\beta-\gamma)\\&+u_{9}\partial^{-1}(u_{1}^{*}u_{8}+u_{4}^{*}u_{9}+u_{10}^{*}u_{11})(\beta-\gamma)+u_{9}^{*}\partial^{-1}(u_{1}u_{8}+u_{4}u_{9}+u_{10}u_{11})(\beta-\gamma),\\
		u_{12t}*=&u_{12x}\gamma^{*}+u_{12x}^{*}\gamma-u_{2}\partial^{-1}(u_{1}u_{12}+u_{2}u_{5}+u_{7}u_{10})(\gamma^{*}-\alpha^{*})-u_{2}\partial^{-1}(u_{1}u_{12}^{*}+u_{2}u_{5}^{*}+u_{7}u_{10}^{*})(\gamma-\alpha)\\&-u_{2}\partial^{-1}(u_{1}^{*}u_{12}+u_{2}^{*}u_{5}+u_{7}^{*}u_{10})(\gamma-\alpha)-u_{2}^{*}\partial^{-1}(u_{1}u_{12}+u_{2}u_{5}+u_{7}u_{10})(\gamma-\alpha)\\&-u_{10}\partial^{-1}(u_{2}u_{7}+u_{3}u_{10}+u_{9}u_{12})(\gamma^{*}-\beta^{*})-u_{10}\partial^{-1}(u_{2}u_{7}^{*}+u_{3}u_{10}^{*}+u_{9}u_{12}^{*})(\gamma-\beta)\\&-u_{10}\partial^{-1}(u_{2}^{*}u_{7}+u_{3}^{*}u_{10}+u_{9}^{*}u_{12})(\gamma-\beta)-u_{10}^{*}\partial^{-1}(u_{2}u_{7}+u_{3}u_{10}+u_{9}u_{12})(\gamma-\beta).
	\end{align*}
	From the two classes of integrable soliton hierarchies constructed above, we select the first hierarchy and employ the Kronecker product to construct its integrable coupling system. Furthermore, we explicitly present the corresponding coupled system for the case $n=1$ as a concrete illustration.

	\section*{Acknowledgment}
	This research was supported by the National Natural Science Foundation of China (No. 11961049, 10601219) and by the Key Project of
	Jiangxi Natural Science Foundation grant (No. 20232ACB201004).
	
	\section*{Declarations} 
	{\bf Conflict of interest} The authors have no conflicts to disclose.\\
	{\bf Ethics approval and consent to participate} All authors approve ethics and consent to participate.\\
	{\bf Consent for publication} All authors consent for publication.\\
	{\bf Data availability} No data was used for the research described in the article.

\end{document}